\documentclass[useAMS,usenatbib,usegraphicx]{mn2e}
\begin{document}
%
%
\title{Molecular Gas and Star Formation in the SAURON Early-type Galaxies}
\author[Combes, Young \& Bureau]{Francoise Combes,$^1$\thanks{E-mail:
francoise.combes@obspm.fr} Lisa M.\ Young,$^{2,3}$ Martin Bureau$^3$\\ 
$^1$Observatoire de Paris, LERMA, 61 Av.\ de l'Observatoire, 75014, Paris,
 France\\
$^2$Physics Department, New Mexico Institute of Mining and Technology,
 Socorro, NM~87801, U.S.A.\\
$^3$Sub-Department of Astrophysics, University of Oxford, Denys
Wilkinson Building, Keble Road, Oxford OX1 3RH}
\maketitle
%
%
\begin{abstract}
We present the results of a survey of CO emission in $43$ of the 48
representative E/S0 galaxies observed in the optical with the {\tt SAURON} 
integral-field spectrograph.  The CO detection rate is 12/43 or $28\%$.
This is lower than previous studies of early-types but can probably be 
attributed to different sample selection criteria.  As expected,
earlier type, more luminous and massive galaxies have a relatively
lower molecular gas content.  We find that CO-rich galaxies tend to have
higher H$\beta$ but lower Fe5015 and Mg$b$ absorption indices than 
CO-poor galaxies.  Those trends appear primarily driven by the age 
of the stars, an hypothesis supported by the fact that the galaxies 
with the strongest evidence of star formation are also the most CO-rich. 
In fact, the early-type galaxies from the current sample appear to
extend the well-known correlations between FIR luminosity, dust mass
and molecular mass of other galaxy types.   The star formation interpretation
is also consistent with the {\tt SAURON} galaxies' radio continuum and FIR
flux ratios, and their inferred star formation efficiencies are similar to
those in spiral galaxies.  It thus appears that we have identified the 
material fueling (residual) star formation in early-type galaxies, and have
demonstrated that it is actively being transformed. Nevertheless, the
lack of strong correlations between the CO content and most stellar
parameters  is compatible with the idea that, in a significant number of
sample galaxies, the molecular gas has been accreted from the
outside and has properties rather independent from the old,
pre-existing stellar component.
\end{abstract}
\begin{keywords}
galaxies: elliptical and lenticular, cD --- galaxies: evolution ---
galaxies: ISM --- galaxies: structure --- Radio lines: galaxies.
\end{keywords}
%
%
\section{INTRODUCTION}
\label{sec:intro}
Over the last twenty years, growing evidence has accumulated that
early-type galaxies (E/S0s) have detectable amounts of cold gas,
sometimes settled in a disc, and recent star formation
\citep*[e.g.][]{bhr92,hrs93,fpmc99,moetal06}. Many E/S0s also show
photometric and kinematic evidence of central stellar discs
\citep*[e.g.][]{b88,nch88}, the presence of which correlates with both
nuclear and global properties
\citep*[e.g.][]{nbs91,fetal97}. Similarly, many E/S0s show evidence
for so-called kinematically decoupled cores (KDCs), i.e.\
kinematically misaligned central components
\citep*[e.g.][]{fih89,eetal04}. These point to the accretion of
external material, question the importance of dissipation, and may
indicate triaxial figures. However, for both central discs and KDCs,
the connection between stellar and gaseous components remains unclear.

Discs in ellipticals can form simultaneously with their host, from the
major mergers of gas-rich galaxies. There has been evidence for this
scenario for a number of years (see \citealt{hggs94} for NGC~7252;
\citealt{gs97} for a review), and recent numerical simulations
illustrate how the disc formation may happen
\citep*[e.g.][]{b02,njb06}. In those mergers, roughly half the gas is
quickly funneled to the center, some causing a burst of star formation
and much settling in a high surface density, mostly molecular central
disc \citep*[e.g.][]{bjc05}. The other half is ejected to large
distances but remains bound to the merger remnant (the new
galaxy). This high angular momentum diffuse gas eventually falls back
and may settle into a more extended, large-scale disc.

Alternatively, some gas could come from stellar mass loss from evolved
stars, and \citet{sw06} recently proposed that essentially all the
molecular gas found in the center of S0s does come from (internal)
stellar mass return, while the atomic gas (HI) mainly originates from
external sources (merger-driven infall, captured dwarfs, regular cold
accretion, etc). Nevertheless, it is as yet not understood why
early-type galaxies only possess $\sim10\%$ of the gas expected from
internal mass loss \citep[e.g.][]{cepr91}.

There have been several CO emission surveys of early-type galaxies
(e.g.\ \citealt{lkrp91}; \citealt*{wch95}; \citealt*{swy06}; see also
the compilation by \citealt*{bgg03}), revealing that early-types are
not always devoid of gas, and can even show a substantial amount of CO
emission. \citet{y02,y05} mapped CO emission in seven Es, revealing
regularly rotating molecular discs in all of them. She showed that
these gas discs should be forming stellar discs with approximately the
same size and mass as typically observed in Es
\citep[e.g.][]{ss96,sbwcm98,eetal04}. Unfortunately, at the present,
we know of no E/S0 galaxy with appropriate data to establish a direct
causal link between the molecular gas and the stellar disc (or
KDC). Integral-field spectroscopy is by far the best tool to study
those decoupled components, but there is simply no overlap between the
existing CO and integral-field samples. The recently completed {\tt
SAURON} survey, however, offers a unique opportunity to reverse this
trend.

Using a custom-designed panoramic integral-field spectrograph
\citep{betal01}, the {\tt SAURON} team have mapped the
stellar kinematics of a large representative sample of early-type
galaxies in the North (see \citealt{zetal02} for the sample definition
and preliminary results). Both the quantity and quality of these data
represent tremendous improvements over previously available
observations of E/S0s. Contrary to expectations, most galaxies show
evidence of kinematic subcomponents which can be uniquely identified
and characterised with the new integral-field maps.  Kinematic 
misalignments, twists, dynamically cold subcomponents, and KDCs are all common  
\citep{eetal04}.   Roughly half of {\tt
SAURON} E/S0s show kinematic evidence of a central disc.

In addition to the stellar distribution and kinematics ($I$, $V$,
$\sigma$, $h_3$ and $h_4$; \citealt{eetal04}), {\tt SAURON}
simultaneously maps the ionised-gas distribution and kinematics
(H$\beta$, [O III] and [N I] lines; \citealt{setal06}), as well as the
age, metallicity and overabundance of the stellar populations (through
the H$\beta$, Mg$b$, Fe5015 and Fe5270 line-strength indices;
\citealt{ketal06}). The central stellar discs and KDCs are often
accompanied by ionised-gas discs, but not always. Similarly, while
most appear old, not all are.  This wealth of information is, however,
not yet assembled into a complete picture of the current state and
formation history of these discs and KDCs.

To better understand the varied phenomena seen in the stellar and
ionised-gas data, a detailed comparison with molecular gas is
warranted. As only a few {\tt SAURON} early-type galaxies had a CO
detection, we recently carried out a single-dish CO survey of most
{\tt SAURON} E/S0s using the IRAM 30~m telescope. The results are
presented and discussed here. Section~\ref{sec:data} presents the
details of our observations as well as literature data, while
Section~\ref{sec:results} summarises the basic results. Correlations
with various physical quantities derived from {\tt SAURON} and other
sources are presented in Section~\ref{sec:correl}, and the far
infrared (FIR)--radio correlation is discussed in
Section~\ref{sec:fir-radio}. We summarise our results and discuss
their implications briefly in Section~\ref{sec:summary}.

%
%
\begin{figure}
\begin{center}
\includegraphics[width=8cm]{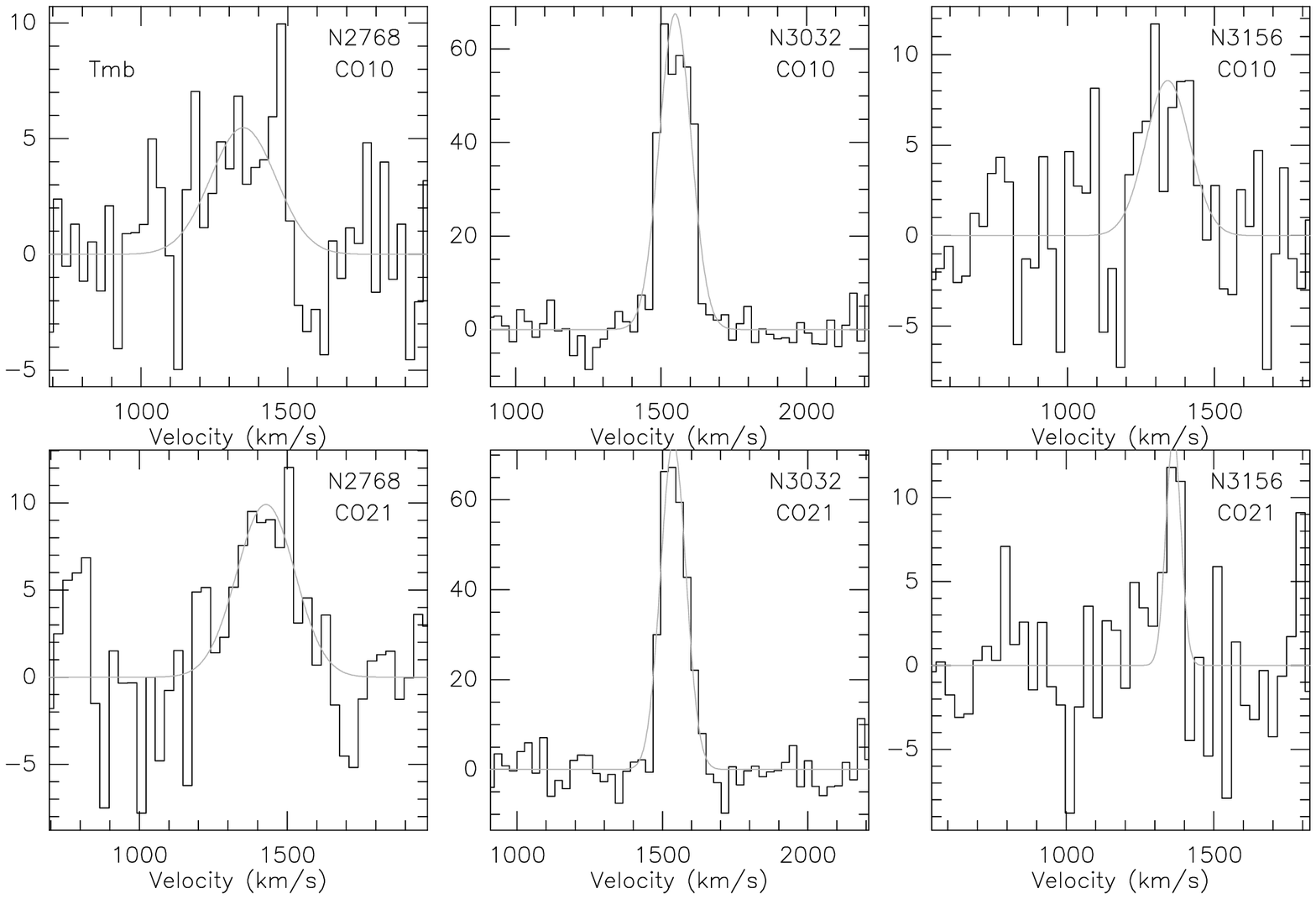}\\
\vspace*{8mm}
\includegraphics[width=8cm]{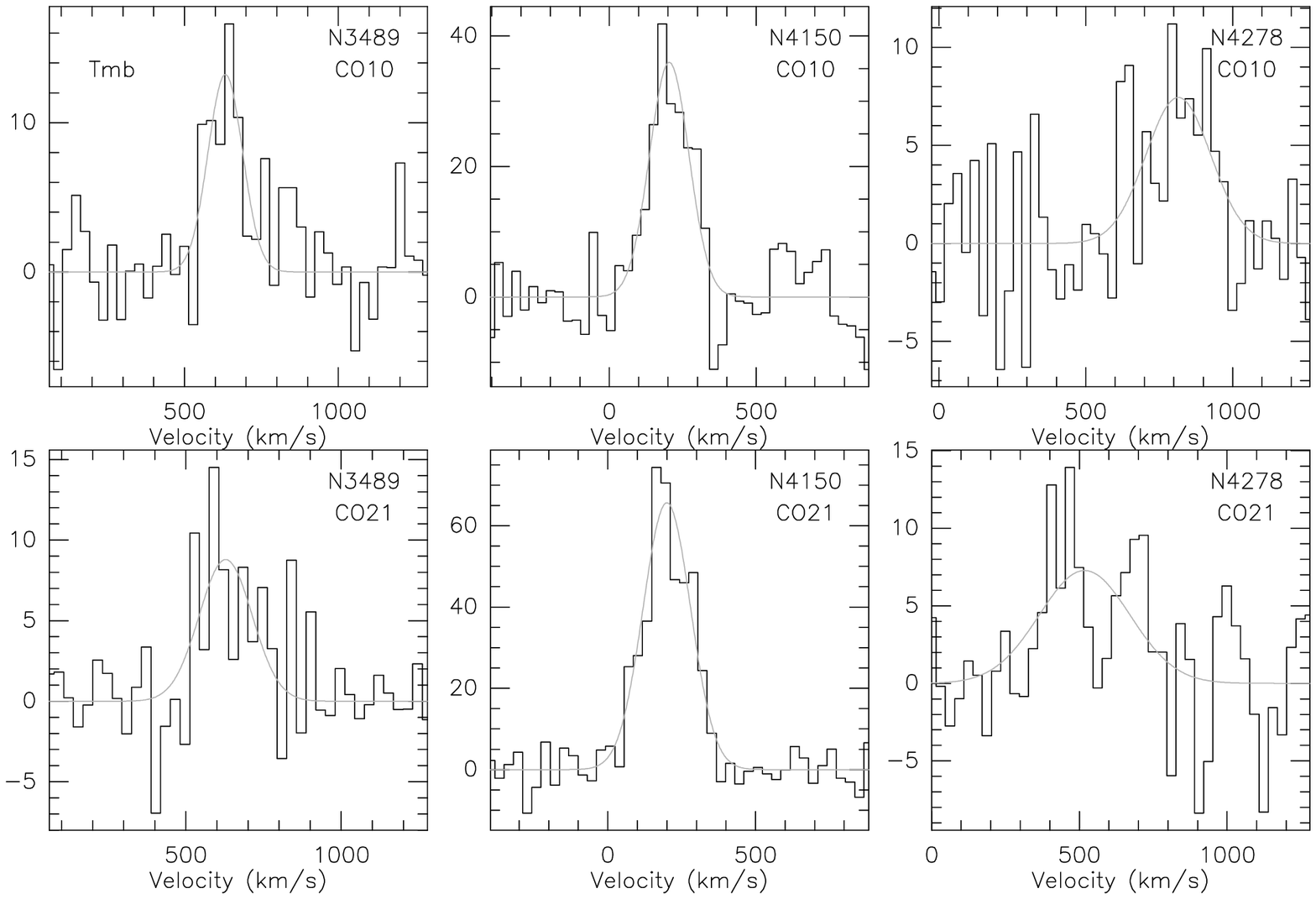}\\
\vspace*{8mm}
\includegraphics[width=8cm]{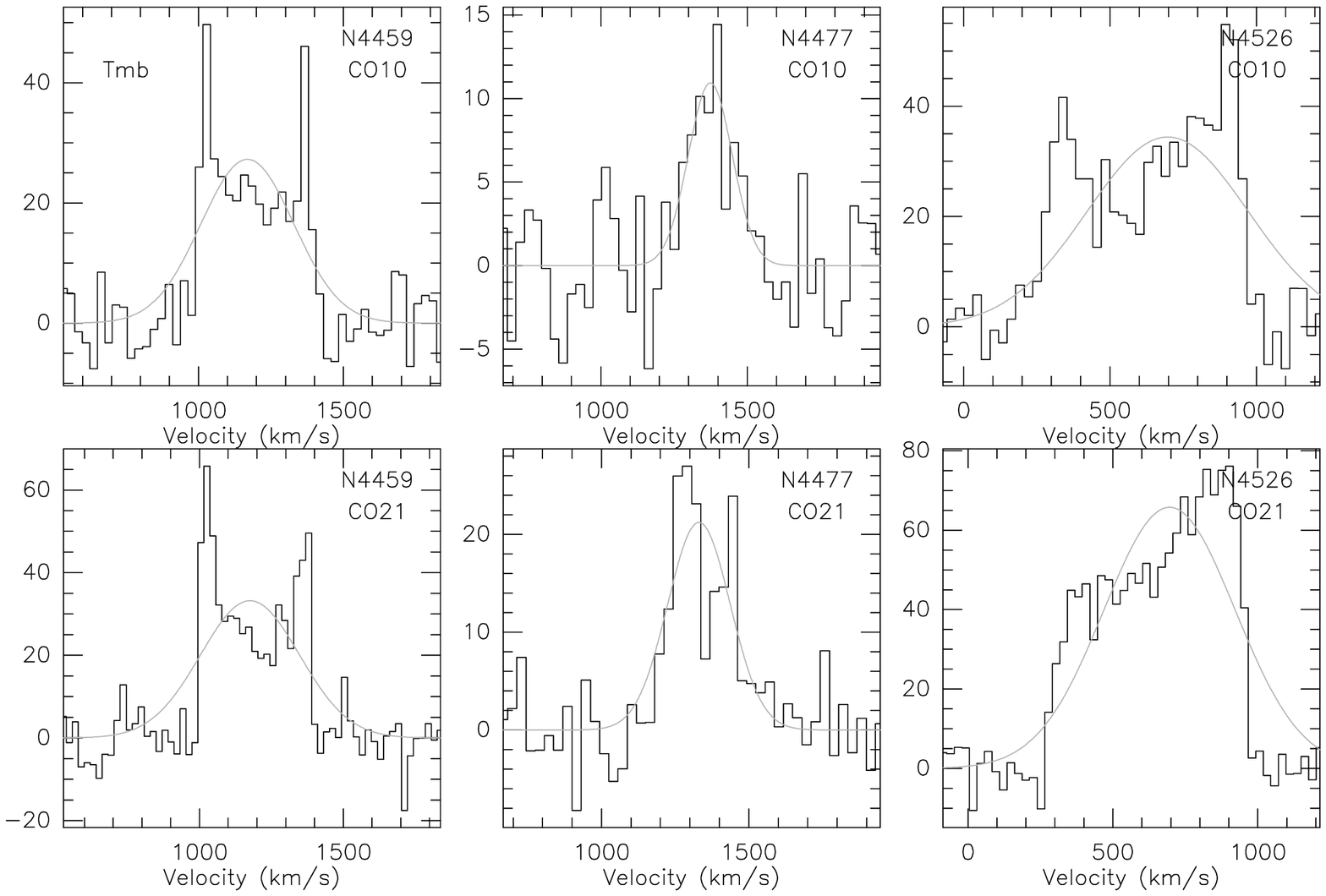}
\caption{CO(1-0) and CO(2-1) IRAM 30~m spectra of the galaxies
detected. The spectra have been binned to $30$~km~s$^{-1}$ and the
scale is $T_{\rm mb}$ in mK. Gaussian fits are overlaid (see
\S~\ref{sec:obs}).}
\label{fig:spectra}
\end{center}
\end{figure}
%
%
\section{DATA}
\label{sec:data}
%
%
\subsection{IRAM 30~m observations and data reduction}
\label{sec:obs}
We simultaneously observed the CO(1-0) and CO(2-1) lines in the centre
of the $43$ early-type galaxies listed in
Tables~\ref{tab:detections}--\ref{tab:upper}. Thirty-nine are from the
{\tt SAURON} representative sample of $48$ E/S0s \citep{zetal02}, and
four were observed with {\tt SAURON} for other purposes
\citep[e.g.][]{detal01,sepb04}. We used the IRAM 30~m telescope at
Pico Veleta, Spain, during April 2003 and April 2004. The beam FWHM is
respectively $23\arcsec$ and $12\arcsec$ at the two frequencies. The
SIS receivers were tuned in single side band mode to the redshifted
frequencies of CO(1-0) at $2.6$~mm and CO(2-1) at $1.3$~mm. The
observations were carried out in wobbler switching, with reference
positions offset by $4\arcmin$ in azimuth. We used the $1$~MHz
back-ends with an effective total bandwidth of $512$~MHz at $2.6$~mm
and the $4$~MHz filterbanks with an effective total bandwidth of
$1024$~MHz at $1.3$~mm.

We spent about $1$--$2$ hours on each galaxy, resulting in a
relatively homogeneous noise level of about $2$--$3$~mK per (binned)
$30$~km~s$^{-1}$ channel for all sources. The system temperatures
ranged between $120$ and $250$~K at $2.6$~mm, and between $300$ and
$500$~K at $1.3$~mm. The pointing was checked every $2$ hours on a
nearby planet or bright quasar. The temperature scale used is in main
beam temperature $T_{\rm mb}$. This was derived from antenna
temperatures $T_{\rm a}^*$ by dividing by $\eta=B_{\rm eff}/F_{\rm
eff}=0.79$ at $2.6$~mm and $0.57$ at $1.3$~mm, where $B_{\rm eff}$ and
$F_{\rm eff}$ are respectively the beam and forward efficiencies.

Each spectrum was summed and reduced using linear baselines, and
binned to $30$~km~s$^{-1}$ channels. Gaussian fits then yielded the
peak intensities, central velocities, velocity FWHMs and integrated
fluxes of the detected lines, listed in Table~\ref{tab:detections}. Some of
the lines are clearly not Gaussian in shape,
but we have checked that integrating the
area under the profiles yields consistent results within the
uncertainties.  The rms noise levels ($1\sigma$ after binning to
$30$~km~s$^{-1}$) are listed for the non-detected galaxies in Table~\ref{tab:upper}.
%
%

\begin{table*}
\caption{Parameters of the 9 early-type galaxies detected in CO.}
\label{tab:detections}
\begin{tabular}{lrrrrrrrrr}
\hline
Galaxy & $I_{\rm (1-0)}$ & $I_{\rm (2-1)}$ & $V_{\rm (1-0)}$ & $V_{\rm (2-1)}$ & $\Delta V_{\rm (1-0)}$ & $\Delta V_{\rm (2-1)}$ & Ratio & $T_{\rm mb\,(1-0)}$ & $T_{\rm mb\,(2-1)}$\\
       & (K km s$^{-1}$) & (K km s$^{-1}$) & (km s$^{-1}$)   & (km s$^{-1}$)   & (km s$^{-1}$)          & (km s$^{-1}$)          &       &  (mK)               & (mK)\\
(1)    & (2)             & (3)             & (4)             & (5)             & (6)                    & (7)                    & (8)   &  (9)                & (10)\\
\hline
NGC2768 &  1.5 (0.28) &  2.5 (0.47) & 1421 (17)           & 1428 (23)           &  221 (51)           &  232 (49)           & 1.6 (0.43) &   6.3 (2.5) &  8.8 (3.5) \\
NGC3032 &  9.3 (0.30) &  7.9 (0.23) & 1549 (\phantom{0}2) & 1537 (\phantom{0}2) &  129 (\phantom{0}4) &  103 (\phantom{0}3) & 0.9 (0.04) &  67.1 (5.1) & 73.7 (5.3) \\
NGC3156 &  1.6 (0.37) &  0.9 (0.23) & 1341 (22)           & 1363 (10)           &  177 (35)           &   60 (18)           & 0.5 (0.19) &   8.9 (3.8) & 14.0 (3.5) \\
NGC3489 &  1.9 (0.30) &  1.9 (0.33) &  633 (10)           &  628 (21)           &  133 (28)           &  201 (45)           & 1.0 (0.24) &  12.7 (2.5) &  8.8 (1.8) \\
NGC4150 &  6.1 (0.49) & 13.2 (0.47) &  204 (\phantom{0}7) &  200 (\phantom{0}3) &  158 (14)           &  189 (\phantom{0}8) & 2.2 (0.20) &  35.4 (5.1) & 64.9 (3.5) \\
NGC4278 &  2.1 (0.41) &  2.8 (0.60) &  815 (27)           &  518 (41)           &  260 (59)           &  361 (76)           & 1.4 (0.39) &   7.6 (2.5) &  7.0 (3.5) \\
NGC4459 & 10.9 (0.48) & 14.3 (0.51) & 1169 (\phantom{0}9) & 1176 (\phantom{0}8) &  377 (16)           &  405 (15)           & 1.3 (0.07) &  26.6 (8.9) & 33.3 (7.0) \\
NGC4477 &  2.1 (0.33) &  5.7 (0.47) & 1374 (14)           & 1331 (11)           &  177 (31)           &  250 (23)           & 2.8 (0.50) &  11.4 (2.5) & 21.1 (5.3) \\
NGC4526 & 23.8 (0.76) & 37.4 (0.72) &  697 (\phantom{0}2) &  695 (\phantom{0}5) &  650 (23)           &  533 (11)           & 1.6 (0.06) &  34.2 (3.8) & 64.9 (8.8) \\
\hline
\end{tabular}
Columns: (1) Source name; (2)--(3) Integrated CO(1-0) and CO(2-1)
emission; (4)--(5) Central velocity in CO(1-0) and CO(2-1); (6)--(7)
Profile FWHM in CO(1-0) and CO(2-1); (8) Integrated CO(2-1)/CO(1-0)
emission ratio; (9)--(10) Main beam peak temperature in CO(1-0) and
CO(2-1). Uncertainties in each quantity are quoted in parenthesis. All
galaxies are part of the {\tt SAURON} representative sample
\citep{zetal02}.
\end{table*}
The total molecular hydrogen (H$_2$) mass was then computed, assuming
the standard conversion ratio
N(H$_2$)/I(CO)~$=3\times10^{20}$~cm$^{-2}$
(K~km~s$^{-1}$)$^{-1}$. Only the CO(1-0) line was used, and the H$_2$
mass corresponds to the $23\arcsec$ beam, equivalent to $2.3$~kpc at
the average distance of our galaxies, $<{\rm D}>\,=20$~Mpc. The H$_2$
masses may thus be underestimated if the molecular gas is more
extended, but the beam does cover roughly one third of the {\tt
SAURON} field-of-view ($33\arcsec\times41\arcsec$ in low-resolution
mode) and comparisons with optically determined parameters are
justified on that basis. Furthermore, experience with interferometric
imaging \citep[e.g.][]{y02,y05} suggests that little flux will be missed
in most cases.

%
%
\subsection{Comparison with previous observations}
\label{sec:comparisons}
Among our $9$ detections, some were already known. We confirm them
here and in most cases improve the signal-to-noise ratio. NGC2768 was
detected by \citet{wch95} with the IRAM 30~m, but with twice lower
integrated intensities and line-widths, probably due to a pointing
problem. NGC3032 was detected by several teams
\citep{sw89,ttkgmty89,kr96} and our results are consistent with
them. NGC3489 and NGC4150 were recently detected by \citet{ws03} with
the IRAM 30~m. They find a wider profile for NGC3489, but with a total
bandwidth which is twice smaller, while NGC4150 is consistent. NGC4459
and NGC4526 were detected with smaller signal-to-noise ratios with the
Kitt Peak 12~m telescope by \citet{sw89}. Their larger beam suggests
that the CO emission is not extended much beyond our own $23\arcsec$
beam. NGC3156, NGC4278 and NGC4477 are detected here for the first
time.

Most of our upper limits are consistent with previous ones but improve
them. \citet{ws03} claim to detect NGC3384 in the CO(2-1) line with
the IRAM 30~m, but nothing is detected in CO(1-0) either with IRAM or
the Kitt Peak 12m. They however only have a narrow bandwidth of
$650$~km~s$^{-1}$ at $1.3$~mm, for a claimed line of zero-power width
of $400$~km~s$^{-1}$, and there is the possibility of a baseline
problem. With a bandwidth twice as large, our CO(2-1) spectrum is not
consistent with their detection, and we consider NGC3384 as an upper
limit in the rest of this paper.
%
%
\subsection{Additional literature data}
\label{sec:literature}
The representative {\tt SAURON} sample of E/S0s is composed of the
$48$ galaxies listed in Table~\ref{tab:correlations} \citep[see
also][]{zetal02}. As stated above, we have observed with IRAM $43$
galaxies previously observed with {\tt SAURON}, of which $39$ belong
to this representative sample. We failed to observe the remaining $9$
because of bad weather, but some have measurements in the
literature. Once rescaled to our own conversion factor and distance
(see \citealt{zetal02} for the latter), \citet*{wgb94} quote an H$_2$
mass of $5.2\times10^8$~$M_{\sun}$ for NGC2685, while \citet{ss02}
report $2.7\times10^7$~$M_{\sun}$. We adopt the more recent
value. \citet{wh01} quote a molecular gas mass of
$1.6\times10^7$~$M_{\sun}$ for NGC4550, and \citet{ws03}
$4.4\times10^6$~$M_{\sun}$ for NGC7457. \citet{ws03} also provide an
upper limit of $1.8\times10^8$~$M_{\sun}$ for NGC7332.

%
%

\begin{table}
\caption{CO upper limits of the 34 non-detected early-type galaxies.}
\label{tab:upper}
\begin{tabular}{lrrrrr}
\hline
Galaxy & $T_{\rm mb\,(1-0)}$ & $T_{\rm mb\,(2-1)}$ & D & 
$\theta_{(1-0)}$ & $\theta_{(2-1)}$ \\
       & (mK)                & (mK) &             (Mpc)  &     (kpc)   &       (kpc) \\
(1)    & (2)                 & (3) &          (4)         &   (5)      &      (6)      \\
\hline
NGC0474           & 2.5 &  3.5 & 31.62  & 3.53  & 1.84\\
NGC1023           & 2.5 &  3.5 & 10.28  & 1.15  & 0.6\\
NGC2549           & 3.8 &  5.3 & 16.75  & 1.87  & 0.97\\
NGC2679$^{\rm a}$ & 3.8 &  3.5 & 30.62  & 3.42  & 1.78\\ 
NGC2695           & 5.1 &  7.0 & 23.23  & 2.59  & 1.35\\
NGC2699           & 2.5 &  3.5 & 23.23  & 2.59  & 1.35\\
NGC2974           & 5.1 &  7.0 & 24.32  & 2.71  & 1.42\\
NGC3377           & 3.8 &  5.3 & 10.67  & 1.19  & 0.62 \\
NGC3379           & 2.5 &  3.5 & 10.67  & 1.19  & 0.62\\
NGC3384           & 5.1 &  7.0 & 10.67  & 1.19  & 0.62\\
NGC3414           & 1.3 &  1.8 & 20.14  & 2.25  & 1.17\\
NGC3608           & 3.8 &  3.5 & 15.56  & 1.74  & 0.91\\
NGC4261$^{\rm a}$ & 2.5 &  3.5 & 31.62  & 3.53  & 1.84\\
NGC4262           & 2.5 &  3.5 & 16.29  & 1.82  & 0.95\\
NGC4270           & 2.5 &  1.8 & 16.29  & 1.82  & 0.95\\
NGC4365$^{\rm a}$ & 3.8 &  3.5 & 17.95  & 2.00  & 1.04\\
NGC4374           & 2.5 &  3.5 & 16.29  & 1.82  & 0.95\\
NGC4382           & 3.8 &  5.3 & 16.29  & 1.82  & 0.95\\
NGC4387           & 3.8 &  3.5 & 16.29  & 1.82  & 0.95\\
NGC4458           & 3.8 &  5.3 & 16.29  & 1.82  & 0.95\\
NGC4473           & 2.5 &  1.8 & 16.29  & 1.82  & 0.95\\
NGC4486           & 2.5 &  1.8 & 16.29  & 1.82  & 0.95\\
NGC4546           & 2.5 &  3.5 & 16.29  & 1.82  & 0.95\\
NGC4552           & 3.8 &  3.5 & 16.29  & 1.82  & 0.95\\
NGC4570           & 5.1 &  5.3 & 16.29  & 1.82  & 0.95\\
NGC5198           & 2.5 &  1.8 & 36.31  & 4.05  & 2.11\\
NGC5308           & 3.8 &  7.0 & 28.31  & 3.16  & 1.65\\
NGC5813           & 2.5 &  5.3 & 26.30  & 2.93  & 1.53\\
NGC5831           & 5.1 &  5.3 & 22.80  & 2.54  & 1.33\\
NGC5838           & 3.8 &  5.3 & 18.71  & 2.09  & 1.09\\
NGC5845           & 2.5 &  1.8 & 21.78  & 2.43  & 1.27\\
NGC5846           & 5.1 &  5.3 & 24.89  & 2.78  & 1.45\\
NGC5982           & 1.3 &  1.8 & 41.88  & 4.76  & 2.44\\
NGC6548$^{\rm a}$ & 3.8 & 10.5 & 33.57  & 3.75  & 1.95\\
\hline
\end{tabular}
\\
Columns: (1) Source name; (2)-(3) upper limits ($1\sigma$) in main
beam temperature for the CO(1-0) and CO(2-1) lines;
(4) adopted distance; (5)-(6) diameters of the CO(1-0) and CO(2-1) beams, in kpc.\\
Notes: a) galaxies observed by {\tt SAURON} but not part of the
representative sample of $48$ E/S0s \citep[see][]{zetal02}.
\end{table}

%
%
\section{BASIC RESULTS}
\label{sec:results}
Figure~\ref{fig:spectra} shows the 30~m spectra obtained in the
CO(1-0) and CO(2-1) lines towards the $9$ galaxies detected. The
derived H$_2$ masses are listed in Table~\ref{tab:correlations},
together with relevant physical parameters against which the molecular
content will be compared.  For the non-detected galaxies, the quoted
mass limits correspond to three times the statistical uncertainty in
an integral over a 300 km~s$^{-1}$ line width.

%
%
\subsection{Excitation of molecular gas: the CO(2-1) to CO(1-0) ratio}
\label{sec:excitation}
The integrated CO(2-1)/CO(1-0) emission ratio is listed in
Table~\ref{tab:detections} for all detected galaxies, as a first
indicator of the excitation temperature of the gas. Since only one
beam has been observed, it is not possible to compare the two lines
with the same spatial resolution over the same region. The ratio is
however generally quite low, $\approx1.4$ on average. In the case of a
concentrated source, with the typical dense and optically thick
molecular gas generally found in galaxy nuclei \citep{bc92}, we would
expect a ratio of up to $4$, the ratio of the beam areas for the two
lines. The lower ratios observed suggest that the CO molecules are
sub-thermally excited, and that the molecular gas density is
relatively low on average.
%
%

\begin{figure*}
\begin{center}
\includegraphics[angle=-90,width=16cm]{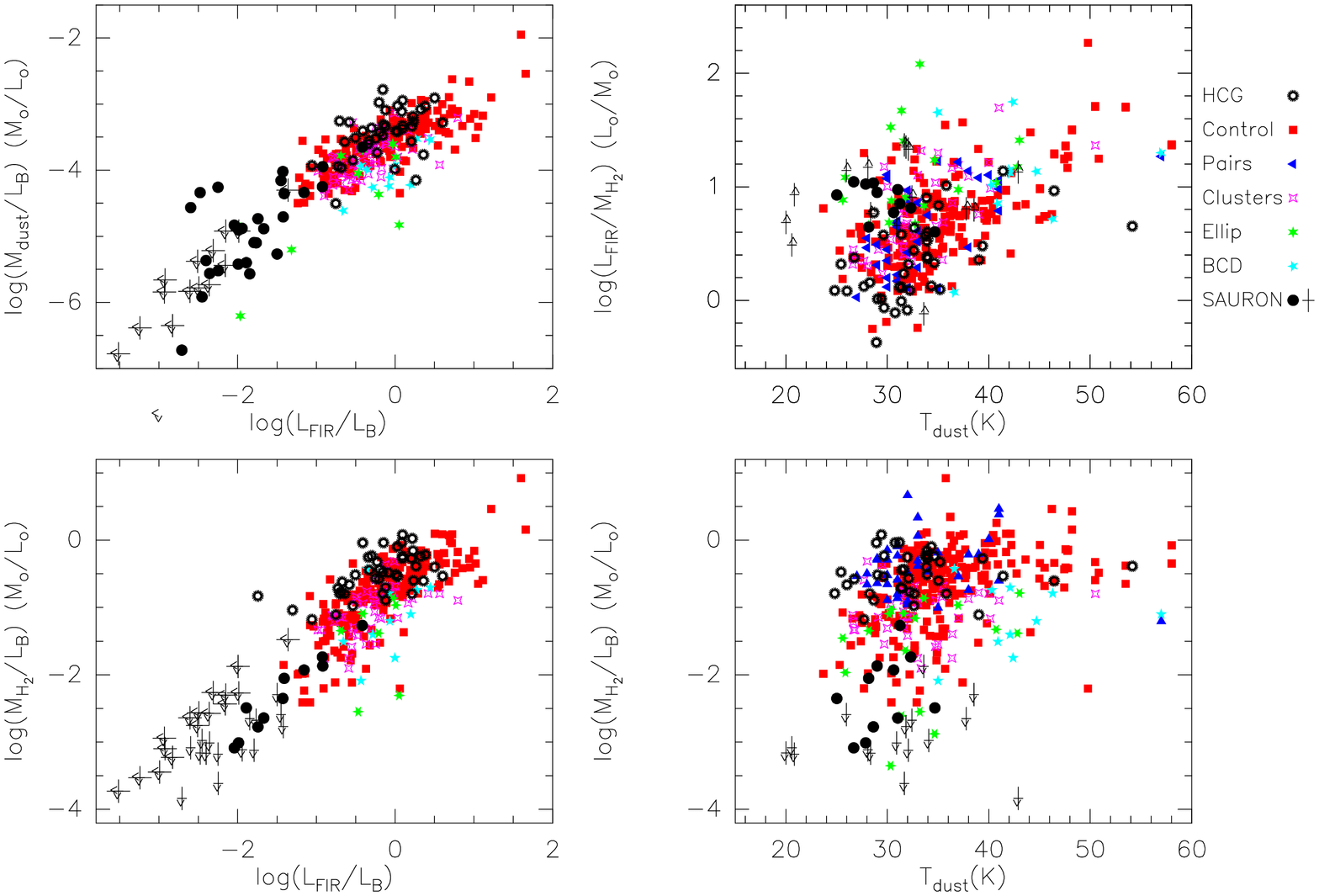}
\caption{Comparison of the {\tt SAURON} representative sample and
other samples of galaxies observed in the CO line. The other samples
are: HCG: Hickson Compact Group galaxies from \citet{lcm98}; Control:
normal galaxies from \citet{yxkr89,yaklr96}, \citet{ss88},
\citet{tsss90} and \citet{s93} (see \citealt{lcm98} for their
definition); Pairs: binary galaxies \citep{cprs94} and starbursts
\citep*{sss91}; Clusters: galaxies in clusters \citep{cbcd91,yxkr89};
Ellip: elliptical galaxies from \citet{wch95}; BCD: blue compact dwarf
galaxies from \citet{itb95} and \citet{sslh92}; SAURON: current sample
\citep[see][]{zetal02}. {\bf Top left:} Dust mass computed from IRAS
fluxes (see text) versus FIR luminosity, both normalised by the blue
luminosity. 
{\bf Top right:} FIR luminosity to
molecular mass ($M_{{\rm H}_2}$) ratio, often used as a tracer of the
star formation efficiency, as a function of dust temperature. Only
{\it IRAS}-detected galaxies are plotted. 
{\bf Bottom
left:} $M_{{\rm H}_2}$ versus FIR luminosity, both normalised by the
blue luminosity. 
{\bf Bottom right:} $M_{{\rm H}_2}$ to blue
luminosity ratio versus dust temperature. Only galaxies detected by
{\it IRAS} are plotted. 
}
\label{fig:sauron-all}
\end{center}
\end{figure*}
%
%
\subsection{CO detection rate}
\label{sec:incidence}
In the {\tt SAURON} representative sample of $48$ E/S0 galaxies, $43$
galaxies have now been searched for CO emission.
We detected $9$ in
both CO(1-0) and CO(2-1) emission and three have previous detections from the
literature, corresponding to a detection rate of
$28\%$.   
By construction, our sample
contains half lenticular and half elliptical galaxies. 
Of the galaxies classed as ellipticals the detection rate is
2/20 ($10\%$) and for lenticulars it is 10/23 ($43\%$).

How does that compare to previous surveys? 
For lenticulars,
the best comparison is with the survey of \citet{ws03}, which was a
volume limited sample of non-Virgo cluster galaxies out to
$20$~Mpc. Their detection rate was $78\%$, much higher than ours. The
difference may however lie in the fact that a volume-limited survey
contains comparatively more small (and low-luminosity) galaxies, which
tend to be richer in molecular gas \citep{lkrp91,swy06}. The {\tt SAURON} sample was selected to
uniformly populate planes of absolute magnitude $M_B$ versus
ellipticity $\epsilon$, so its distribution in magnitude is
approximately flat by construction, while the magnitude distribution
of the complete sample from which it was drawn is much more skewed
toward low-luminosity objects (see Fig.~1 in \citealt{zetal02}). 
The median absolute magnitude of the {\tt SAURON} sample is $M_B =
-19.9$ whereas for the sample in \citet{ws03} it is $M_B = -18.8$.  The
detection rate of \citet{ws03} drops from $78\%$ for the entire sample 
to $64\%$ for galaxies
brighter than $M_B = -18.8$, so our lower detection rate is roughly 
consistent with this trend. 

There are probably several factors influencing the detection rates in
the ellipticals.
\citet{kr96},
\citet{wch95} and \citet{lkrp91} got detection rates of $30$--$40\%$
for FIR-selected samples. Our detection rate is lower, most
likely because the {\tt SAURON} galaxies are not
FIR-selected. However, we also have a lower detection rate
than \citeauthor{swy06} (\citeyear{swy06}; $33\%$), who studied a
volume-limited sample of ellipticals. 
The median $M_B$ of their sample, $-19.5$, is also a bit fainter than
the median of the {\tt SAURON} sample.
Environment may
also play a role, since again by construction half of the {\tt SAURON}
galaxies are in clusters. The effects of environment are however harder to gauge, since truly
isolated early-type galaxies are rare, and we found no correlation
between molecular gas content and environment (local galaxy number
density; \citealt{t88}).   Similarly, \citet{ky89} found no deficiency
in the molecular gas content of Virgo cluster spirals even though they
are strongly deficient in atomic gas.
At the moment, it nevertheless seems entirely
plausible that differences between our results and previous works can
be attributed mainly to different sample selection criteria.
%
%
\subsection{Star formation tracers}
\label{sec:sf_tracers}
In spiral galaxies the FIR emission is usually assumed to trace star
formation activity.  Therefore it is of interest to compare the FIR and
molecular properties of the {\tt SAURON} galaxies to those of star forming
galaxies.
A ``bolometric'' FIR flux is 
derived for each galaxy from the {\it IRAS} $60$ and $100$~$\micron$ fluxes,
\begin{equation}
FIR \equiv1.26\times10^{-14}~{\rm W}~{\rm m}^{-2}\,
                 (2.58~S_{60\micron} + S_{100\micron})\,,
\end{equation}
where $S_{60\micron}$ and $S_{100\micron}$ are measured in Jy. 
The total FIR luminosity is then $L_{\rm FIR} = (FIR)\, 4\pi D^2$.

From the ratio of the {\it IRAS} $60$ and $100$~$\micron$ fluxes, we
have derived dust temperatures assuming $\kappa_\nu\propto\nu$, where
$\kappa_\nu$ is the mass opacity of the dust at frequency $\nu$. The
average dust temperature for the {\it IRAS}-detected {\tt SAURON}
galaxies is $30\pm5$~K. This is relatively low, indicating a low star
formation density. By comparison, starburst galaxies have higher dust
temperatures, $\approx40$~K or more \citep[e.g.][]{sm96}.

Knowing the dust temperature $T_{\rm dust}$ and the $100$~$\micron$
flux $S_{100}$, we can derive the dust mass as
\begin{eqnarray*}
 M_{\rm dust} & = & 4.8\times10^{-11}\,{S_\nu\,d_{\rm Mpc}^{\,2}
 \over \kappa_\nu\,B_\nu(T_{\rm dust})}\ M_{\sun} \\
              & = & 5\,S_{100\micron}\,d_{\rm Mpc}^{\,2}\,
  \left\{\exp(144/T_{\rm dust}) - 1 \right\}\ M_{\sun},
\end{eqnarray*}
where $S_\nu$ is the FIR flux measured in Jy, $d_{\rm Mpc}$ is the
distance in Mpc, $B_\nu$ is Planck's function, and we use a mass
opacity coefficient of $25$~cm$^{2}$~g$^{-1}$ at $100$~$\micron$
\citep{h83}.
The presence of colder ($T \la 20$~K) dust could significantly change the derived dust
masses but such an analysis would require additional millimeter and
submillimeter continuum observations \citep[e.g.][]{lsrhk04}.
For the current work such cold dust is neglected.

Figure \ref{fig:sauron-all} compares the dust and molecular content of
the {\tt SAURON} representative sample with various other samples
studied in the literature. Those span a wide range of masses,
metallicities and environments (field, group and cluster), with
different star formation rates (starburst, quiescent) and different
morphological types (elliptical, spiral, dwarf). The various samples
are described in \citet*{lcm98} and in Figure~\ref{fig:sauron-all}
itself.

The main correlations in Figure~\ref{fig:sauron-all} are the
well-known relations between the molecular content, dust content and
FIR luminosity. The early-type galaxies from the present sample extend
the relations toward low FIR-to-blue luminosity ratios $L_{\rm FIR}/L_B$. Their molecular
content normalised by the blue luminosity is about $2$ orders of
magnitude lower than in normal spiral galaxies. In previous works,
this dependency with type was not found to be so large, even though
one of the essential characteristics of early-type galaxies is their
low gas fraction \citep[e.g.][]{cbcd91,ys91}. This result was generally
attributed to metallicity effects, which influence the CO-to-H$_2$
conversion factor. 
Indeed, the content in molecular gas could in fact be lower in
early-types, but this would not be noticed in the CO emission (when
using a fixed CO-to-H$_2$ conversion factor) because of the
compensating effect of a larger CO abundance per H$_2$
molecule. Early-type galaxies are in general more
massive and, metallicity increasing monotonically with mass, also have
higher metallicities. For a given decrease in molecular gas content,
the CO emission is thus not dropping as fast. The present sample, due
to the improved sensitivity of the telescope, extends the range in gas
content to lower values, but it is not entirely clear why the
molecular gas content dependence on morphological type is more
acute than has been found in the past. The answer probably lies again in the different criteria used
to select the targets (see \S~\ref{sec:incidence}), but our results
are also suggestive that the metallicity-driven increase in CO
luminosity per unit H$_2$ mass is more
than compensated by the associated decrease in gas content.

The $L_{\rm FIR}/M_{{\rm H}_2}$ ratio is usually called the star
formation efficiency (SFE), and somewhat surprisingly, the {\tt SAURON}
galaxies tend to have SFEs in the range of the highest values for
normal spirals.
The low average dust
temperatures derived above are expected for such low values of the
molecular gas content.
%
%

\begin{figure}
\begin{center}
\includegraphics[angle=-90.0,width=8cm]{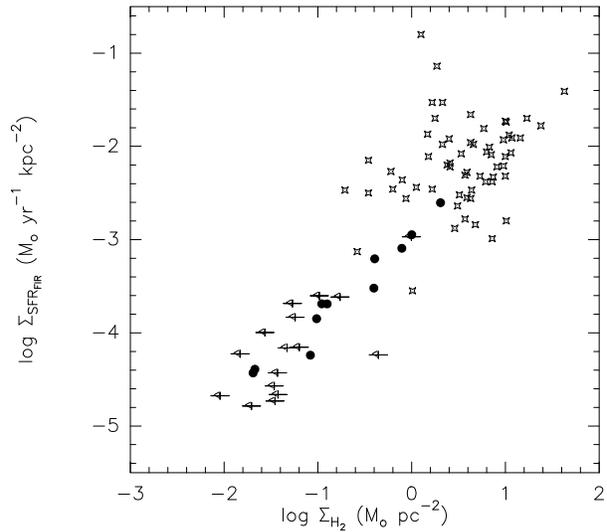}
\caption{Kennicutt-Schmidt diagram of our sample galaxies, showing the
FIR-derived star formation rate surface density as a function of the
surface density of H$_2$ (see text). CO detections are shown with
solid dots and upper limits by arrows. The normal spiral galaxy sample
of \citet{k98} is also plotted for comparison (stars).}
\label{fig:ken_sfr}
\end{center}
\end{figure}
Figure~\ref{fig:ken_sfr} shows the so-called ``Kennicutt-Schmidt
diagram'' \citep[e.g.][]{k98} of our objects. This relates the star
formation rate (SFR) surface density (as derived from the FIR
luminosity and normalised by the area within $R_{25}$, the radius at
the $25$~mag~arcsec$^{-2}$ isophote in $B$) to the surface density of
H$_2$ (the H$_2$ mass detected normalised by the same area). The
normal spiral galaxy sample of \citet{k98} is also plotted for
comparison. Our sample objects appear to follow the same correlation
as normal spirals, but the relation is tighter (at least for the
detections) and extends that for normal spirals by almost $2$ orders
of magnitude down in SFR.
%
%
\section{MOLECULAR MASS CORRELATIONS}
\label{sec:correl}
We have explored many possible correlations of the molecular gas mass
with other galaxy parameters available in catalogues or the
literature, particularly the {\tt SAURON} survey papers. Most
quantities do not correlate significantly with H$_2$ mass, but a few
do show interesting trends and are discussed below for the {\tt
SAURON} representative sample. Table~\ref{tab:correlations} lists all
the parameters used in those correlations.

No correlation was found between $M_{{\rm H}_2}/L_B$ and the galaxy
inclination, the $V/\sigma$ parameter quantifying rotational support
or with any apparent flattening or axis ratio measured at any
radius. Normally, face-on galaxies are much easier to detect in CO
since they have very narrow line widths (all the emission is
concentrated in a few channels), and conversely edge-on galaxies are
harder to detect because of their broad line widths (the emission is
spread out over many channels with low intensity) and possible
confusion with a baseline ripple. However, the absence of a
correlation between gas content and any inclination indicator in the
present study suggests that there is no detection bias related to the
line width.
%
%
\subsection{Correlations with dust content}
\label{sec:dust_corr}
\begin{figure}
\includegraphics[width=8cm]{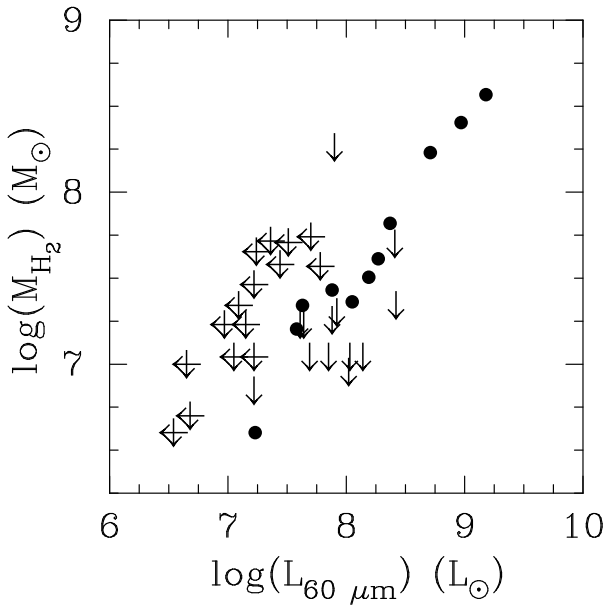}
\caption{Molecular gas mass versus {\it IRAS} $60$~$\micron$ luminosity
  for the {\tt SAURON} representative sample. The correlation is
  excellent. Detections are shown with solid dots and upper limits by
  arrows.}
\label{fig:l60}
\end{figure}

As a reference, we show in Figure~\ref{fig:l60} the relationship
between molecular gas mass and {\it IRAS} $60$~$\micron$ luminosity,
by far the strongest correlation we observe (see also the bottom-left
panel in Fig.~\ref{fig:sauron-all}). All galaxies detected in CO are
in fact also detected by {\it IRAS}. They follow a very tight relation
which, as suggested above, extends to lower H$_2$ masses a similar
correlation for later galaxy types \citep[e.g.][]{ys91}.

As might be expected, the {\tt SAURON} galaxies which are detected in
CO also show dust features in the unsharp-masked optical images of
\citet{setal06}. The one exception to this statement is NGC7457, for
which \citet{ws03} report a small amount of molecular gas and which
shows no small-scale dust feature beyond the nucleus
($R\ga1\arcsec$). The dust morphologies across the sample range from
well developed large disks ($\ga10\arcsec$ diameter; e.g.\ NGC2685,
NGC3032, NGC4459 and NGC4526) to small nuclear features
($\la5\arcsec$; e.g.\ NGC3156, NGC4150 and NGC4477) and even larger
scale irregular patches or lanes (e.g.\ NGC3156, NGC4150, NGC3489 and
NGC4278). 

Dust morphologies are commonly assumed to indicate, at least
qualitatively, how long the cold gas has been present in a galaxy,
since it should take several dynamical times for the gas to settle
into the most regular disks \citep[e.g.][]{ttfcjbr01}. If this
argument holds for the molecular gas, we can infer that the galaxies
with large, regular dust disks have had their cold gas longer than the
galaxies with irregular dust morphologies
\citep[see also][]{setal06}.  Interferometric maps revealing the
kinematics of the molecular gas should give more quantitative insights
into its origin.
%
%

\begin{table*}
\caption{Galaxy parameters for the {\tt SAURON} representative sample.}
\label{tab:correlations}
\begin{tabular}{lrrrrrrrrrrrrr}
\hline
Galaxy & $M_{{\rm H}_2}$     & Type & $M_B$ & $\sigma_{\rm c}$ & $L_{60\micron}$ &
$L_{{\rm H}\alpha}$ & $(B-V)_{\rm e}$ & H$\beta$ & Fe5015 & Mg$b$ &
Age$_{\rm c}$ & $i$   & $V/\sigma$ \\ 
       & ($10^8$ $M_{\sun}$) &      & (mag) & (km s$^{-1}$)    & ($L_{\sun}$)    & 
(erg s$^{-1}$)      & (mag)           & (\AA)    & (\AA)  & (\AA) &
(Gyr)         & (deg) & \\
(1)    & (2)                 & (3)  & (4)   & (5)              & (6)             &
(7)                 & (8)             & (9)      & (10)   & (11)  &
(12)          &  (13) & (14) \\
\hline
NGC 474 & $<$ 0.38 & -2.2 & -20.42 & 170 &$<$7.44 & 39.44 &  0.940 & 1.81 & 4.85 & 3.45 &   -- & 42 &  -- \\
NGC 524 &       -- & -1.5 & -21.40 & 245 &   8.92 & 39.00 &  1.070 & 1.50 & 5.38 & 4.20 &   -- &  9 &  -- \\
NGC 821 &       -- & -4.2 & -20.44 & 208 &$<$7.35 &    -- &  1.020 & 1.57 & 4.65 & 3.72 &   -- & 69 & 0.4 \\
NGC1023 & $<$ 0.04 & -2.6 & -20.42 & 206 &$<$6.54 & 38.46 &  1.010 & 1.51 & 4.98 & 4.19 &  4.7 & 90 & 1.1 \\
NGC2549 & $<$ 0.18 & -2.0 & -19.36 & 146 &   7.88 & 38.80 &  0.955 & 2.02 & 5.10 & 3.54 &  2.5 & 90 & 1.0 \\
NGC2685 &     0.27 & -0.7 & -19.05 &  99 &   7.88 & 39.37 &  0.935 & 2.04 & 4.21 & 3.04 &   -- & 72 & 1.1 \\
NGC2695 & $<$ 0.45 & -2.4 & -19.38 & 220 &$<$7.24 &    -- &  0.980 & 1.27 & 4.34 & 3.94 & 10.2 & 52 & 0.8 \\
NGC2699 & $<$ 0.23 & -5.0 & -18.85 &  -- &     -- & 38.02 &  0.980 & 1.76 & 4.76 & 3.58 &  1.5 & 29 &  -- \\
NGC2768 &     0.41 & -3.1 & -21.15 & 188 &   8.27 & 39.65 &  0.960 & 1.64 & 4.43 & 3.64 &  2.5 & 75 & 0.8 \\
NGC2974 & $<$ 0.50 & -3.6 & -20.32 & 229 &   8.41 & 39.96 &  1.005 & 1.71 & 5.14 & 4.09 &  7.9 & 75 & 0.9 \\
NGC3032 &     2.54 & -1.7 & -18.77 &  82 &   8.97 & 39.42 &  0.630 & 3.98 & 4.02 & 2.14 &  0.9 & 31 &  -- \\
NGC3156 &     0.22 & -2.4 & -18.08 &  84 &   7.63 & 39.18 &  0.770 & 3.00 & 3.66 & 1.79 &   -- & 66 & 0.9 \\
NGC3377 & $<$ 0.07 & -4.1 & -19.24 & 136 &   7.22 & 38.96 &  0.905 & 1.86 & 4.37 & 3.20 &   -- & 90 & 0.7 \\
NGC3379 & $<$ 0.05 & -4.1 & -20.16 & 206 &$<$6.68 & 38.18 &  0.975 & 1.44 & 4.81 & 4.15 & 11.5 & 30 & 0.2 \\
NGC3384 & $<$ 0.10 & -2.6 & -19.56 & 142 &$<$6.65 & 37.92 &  0.955 & 1.84 & 5.12 & 3.70 &  3.2 & 90 & 0.9 \\
NGC3414 & $<$ 0.09 & -2.5 & -19.78 & 246 &   8.02 & 39.72 &  0.930 & 1.50 & 4.49 & 3.72 &  3.6 & 40 & 0.9 \\
NGC3489 &     0.12 & -2.1 & -19.32 & 138 &     -- & 39.64 &  0.845 & 2.60 & 4.40 & 2.60 &  1.7 & 63 & 0.5 \\
NGC3608 & $<$ 0.15 & -4.3 & -19.54 & 204 &     -- & 38.15 &  1.000 & 1.60 & 4.73 & 3.75 &  8.9 & 51 & 0.1 \\
NGC4150 &     0.66 & -2.4 & -18.48 & 148 &   8.37 & 38.91 &  0.830 & 2.87 & 4.11 & 2.35 &  1.5 & 55 &  -- \\
NGC4262 & $<$ 0.11 & -2.6 & -18.88 & 186 &   7.69 & 39.41 &  0.970 & 1.49 & 4.46 & 3.90 & 10.2 & 30 &  -- \\
NGC4270 & $<$ 0.11 & -1.1 & -18.28 & 140 &$<$7.05 &    -- &  0.950 & 1.74 & 4.67 & 3.24 &   -- & 78 & 0.7 \\
NGC4278 &     0.23 & -4.6 & -19.93 & 252 &   8.05 & 39.95 &  0.960 & 1.41 & 4.59 & 4.21 &   -- & 27 & 0.2 \\
NGC4374 & $<$ 0.11 & -3.5 & -21.23 & 297 &   8.14 & 39.30 &  1.000 & 1.42 & 4.66 & 4.08 &   -- & 35 & 0.1 \\
NGC4382 & $<$ 0.17 & -1.8 & -21.28 & 177 &   7.61 &    -- &  0.895 & 1.98 & 4.72 & 3.36 &  1.7 & 45 & 0.4 \\
NGC4387 & $<$ 0.17 & -3.4 & -18.34 & 117 &$<$7.15 &    -- &  0.965 & 1.51 & 4.42 & 3.67 &   -- & 66 & 0.5 \\
NGC4458 & $<$ 0.17 & -3.8 & -18.42 & 102 &$<$6.97 &    -- &  0.915 & 1.59 & 3.97 & 3.27 &   -- & 25 & 0.2 \\
NGC4459 &     1.70 & -2.0 & -19.99 & 174 &   8.71 & 38.89 &  0.970 & 1.85 & 4.68 & 3.43 &  1.9 & 46 & 0.4 \\
NGC4473 & $<$ 0.11 & -4.2 & -20.26 & 191 &$<$7.22 &    -- &  0.990 & 1.50 & 4.88 & 4.03 & 13.2 & 90 & 0.3 \\
NGC4477 &     0.32 & -1.8 & -19.96 & 172 &   8.19 & 39.42 &  0.970 & 1.59 & 4.71 & 3.84 &   -- & 28 & 0.2 \\
NGC4486 & $<$ 0.11 & -4.0 & -21.79 & 351 &   8.03 & 39.70 &  0.980 & 1.15 & 4.70 & 4.62 & 15.1 & 65 & 0.0 \\
NGC4526 &     3.69 & -1.6 & -20.68 & 256 &   9.18 & 39.22 &  0.975 & 1.62 & 4.93 & 4.16 &  2.8 & 90 & 1.0 \\
NGC4546 & $<$ 0.11 & -2.6 & -19.98 & 242 &   7.85 & 39.74 &  0.990 & 1.54 & 4.64 & 3.92 &   -- & 90 & 0.9 \\
NGC4550 &     0.16 & -2.3 & -18.83 &  80 &   7.58 & 39.29 &  0.890 & 1.99 & 4.32 & 2.93 &   -- & 90 & 1.5 \\
NGC4552 & $<$ 0.17 & -3.3 & -20.58 & 264 &   7.64 & 38.69 &  1.000 & 1.37 & 5.33 & 4.55 &  8.9 & 29 & 0.0 \\
NGC4564 &       -- & -4.1 & -19.39 & 168 &     -- &    -- &  0.965 & 1.55 & 4.85 & 4.00 &  3.6 & 90 & 0.9 \\
NGC4570 & $<$ 0.22 & -1.7 & -19.54 & 188 &$<$7.09 & 36.94 &  0.970 & 1.45 & 4.79 & 3.99 &   -- & 90 &  -- \\
NGC4621 &       -- & -4.0 & -20.64 & 245 &$<$7.05 &    -- &  0.975 & 1.43 & 4.75 & 4.14 &  6.9 & 54 & 0.5 \\
NGC4660 &       -- & -4.1 & -19.22 & 191 &$<$7.13 &    -- &  0.990 & 1.47 & 4.80 & 4.07 &   -- & 50 & 0.8 \\
NGC5198 & $<$ 0.55 & -3.4 & -20.38 & 195 &$<$7.70 & 39.13 &  0.985 & 1.56 & 4.70 & 3.93 & 15.1 & 38 & 0.0 \\
NGC5308 & $<$ 0.51 & -1.1 & -20.27 & 260 &$<$7.51 &    -- &  0.925 & 1.46 & 4.92 & 4.23 &   -- & 90 & 0.9 \\
NGC5813 & $<$ 0.29 & -4.5 & -20.99 & 238 &$<$7.22 & 39.36 &  1.010 & 1.52 & 4.74 & 4.12 & 15.1 & 57 & 0.0 \\
NGC5831 & $<$ 0.44 & -4.2 & -19.73 & 175 &     -- & 38.19 &  0.985 & 1.78 & 4.78 & 3.32 &  3.2 & 39 & 0.2 \\
NGC5838 & $<$ 0.22 & -3.0 & -19.87 & 274 &   8.42 & 38.79 &  1.010 & 1.58 & 5.05 & 4.19 &   -- & 90 & 0.8 \\
NGC5845 & $<$ 0.20 & -4.1 & -18.58 & 221 &   7.92 &    -- &  1.120 & 1.52 & 5.38 & 4.41 &  8.9 & 69 & 0.6 \\
NGC5846 & $<$ 0.52 & -4.2 & -21.24 & 250 &$<$7.36 & 39.48 &  1.030 & 1.32 & 4.89 & 4.38 &  7.9 & 25 & 0.0 \\
NGC5982 & $<$ 0.37 & -3.9 & -21.46 & 251 &$<$7.78 & 38.99 &  0.940 & 1.60 & 5.18 & 3.98 & 13.2 & 62 & 0.3 \\
NGC7332 & $<$ 1.82 & -1.7 & -19.93 & 134 &   7.90 & 39.67 &  0.905 & 2.16 & 4.80 & 3.17 &   -- & 90 & 1.0 \\
NGC7457 &     0.04 & -2.2 & -18.81 &  75 &   7.23 & 38.23 &  0.900 & 2.25 & 4.33 & 2.68 &   -- & 67 & 1.0 \\
\hline
\end{tabular}
\\
Columns: (1) NGC number; (2) CO-derived H$_2$ mass (this paper); (3)
Morphological type (HyperLEDA: http://leda.univ-lyon1.fr/;
\citealt{zetal02}); (4) Total absolute blue magnitude \citep{zetal02};
(5) Central stellar velocity dispersion (HyperLEDA;
\citealt{zetal02}); (6) {\it IRAS} $60$~$\micron$ luminosity
(logarithmic; NED: http://nedwww.ipac.caltech.edu/index.html); (7)
Total H$\alpha$ luminosity within the {\tt SAURON} field-of-view,
derived assuming an H$\alpha/$H$\beta$ flux ratio of $2.86$
(logarithmic; \citealt{setal06}); (8) Effective $B-V$ color
(HyperLEDA; \citealt{zetal02}); (9)--(11) Absorption line-strength
indices, luminosity-weighted over an effective radius \citep{ketal06};
(12) luminosity-weighted age of the stellar populations, as derived
from line-strength indices measured in a central
$4\arcsec\times4\arcsec$ aperture assuming a single stellar population
\citep{mcetal06}; (13) Inclination (HyperLEDA); (14) Rotational
support diagnostic $V_{rot}/\sigma$ (HyperLEDA).
\end{table*}
%
%
%
%
\subsection{Correlations with structural parameters}
\label{sec:structure}
\begin{figure*}
\includegraphics[width=16cm]{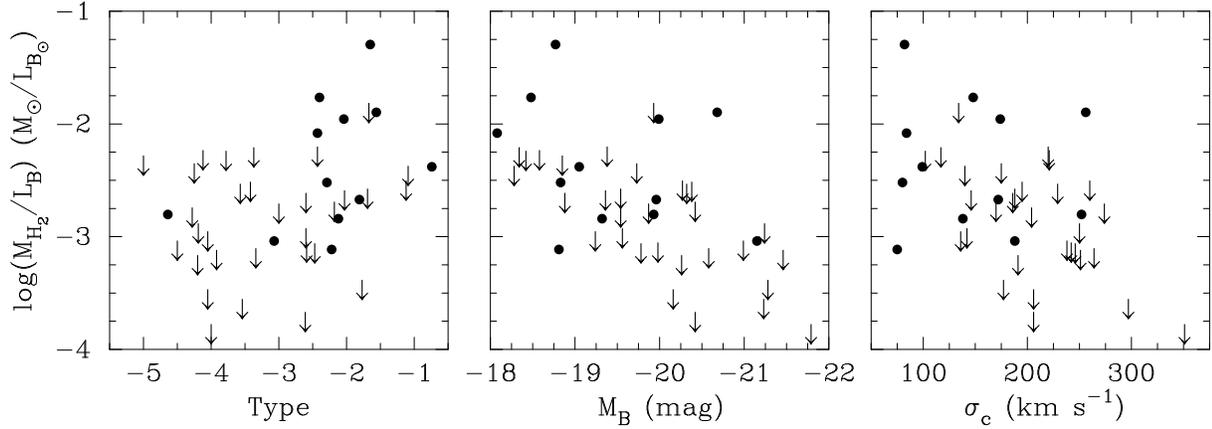}
\caption{Molecular gas content (normalised by the total blue
  luminosity) versus structural parameters for the {\tt SAURON}
  representative sample. {\bf Left:} Morphological type (HyperLEDA;
  \citealt{zetal02}). {\bf Centre:} Total absolute $B$ magnitude
  \citep{zetal02}. {\bf Right:} Central stellar velocity dispersion
  (HyperLEDA). Detections are shown with solid dots and upper limits
  by arrows.}
\label{fig:structure}
\end{figure*}

Even within the narrow morphological range defined by the E/S0s of the
{\tt SAURON} representative sample, Figure~\ref{fig:structure} shows
that earlier type galaxies (which are also generally more luminous and
massive) have relatively less molecular gas per unit luminosity than
later type galaxies. It is also clear that our detections are almost
exclusively confined to the later types, in agreement with the
standard view of early-type galaxies being gas poor with little to no
star formation. Those trends are consistent with previous results
\citep[e.g.][]{lkrp91,swy06} but are not very strong, which could be
due to the compensating effects of metallicity and evolution mentioned
above.
%
%
\subsection{Correlations with stellar populations}
\label{sec:indices}
\begin{figure*}
\includegraphics[width=16cm]{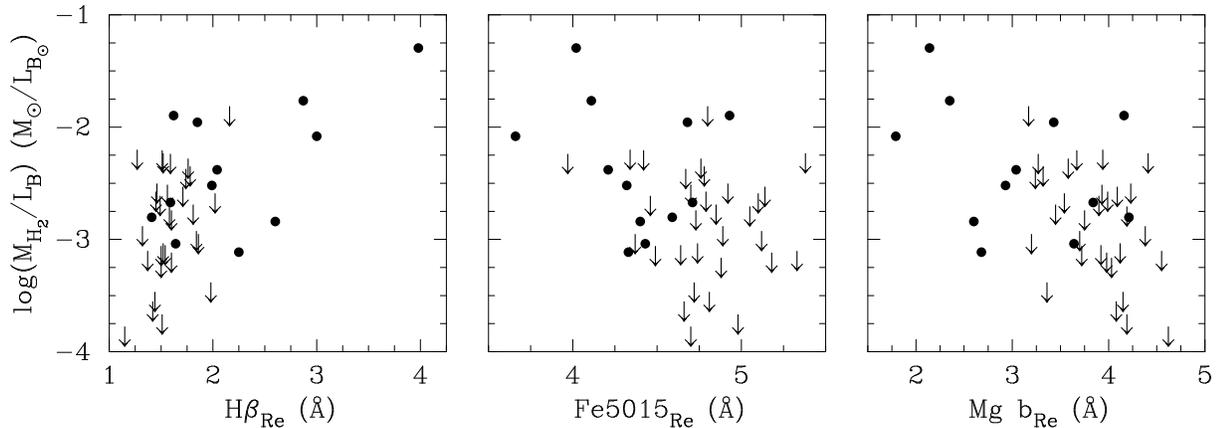}
\caption{Molecular gas content (normalised by the total blue
  luminosity) versus absorption line-strength indices
  (luminosity-weighted within one effective radius) from
  \citet{ketal06} for the {\tt SAURON} representative sample. {\bf
  Left:} H$\beta$ index. {\bf Centre:} Fe51015 index. {\bf Right:}
  Mg$b$ index. Detections are shown with solid dots and upper limits
  by arrows.}
\label{fig:spop}
\end{figure*}

Figure~\ref{fig:spop} shows the absorption line-strength indices of
H$\beta$, Fe5015 and Mg$b$, measured within one effective radius
$R_{\rm e}$ (the radius encompassing half the light) with {\tt SAURON}
\citep[see][]{ketal06}, as a function of the molecular gas content
normalised by the blue luminosity.  There is a clear trend in the sense
that the galaxies with large H$\beta$ line-strengths are more likely to
have detectable CO emission.  Galaxies with smaller Fe5015 and Mg$b$
indices are also more likely to be detected.  
Similar results are obtained when the optical indices are measured
within a smaller aperture ($R_{\rm e}/8$).

H$\beta$ is primarily a tracer of age, especially at high values
(H$\beta\ga2.0$), suggesting that galaxies with younger stellar
populations are relatively richer in molecular gas. This is of course
expected since stars are born within molecular clouds, but to our
knowledge this correlation is observed here for the first time. The
Fe5015 and Mg$b$ indices can respectively be thought of as tracing the
metallicity and $\alpha$-element abundance, which could suggest that
CO-rich galaxies are more metal and $\alpha$-element poor (and thus
may have had a slow star formation fuelled by relatively pristine
gas), but those relationships and inferences are less
straightforward. More direct probes of the star formation and stellar
population parameters are thus preferable.
%
%
\subsection{Correlations with star formation}
\label{sec:sf}
\begin{figure*}
\includegraphics[width=16cm]{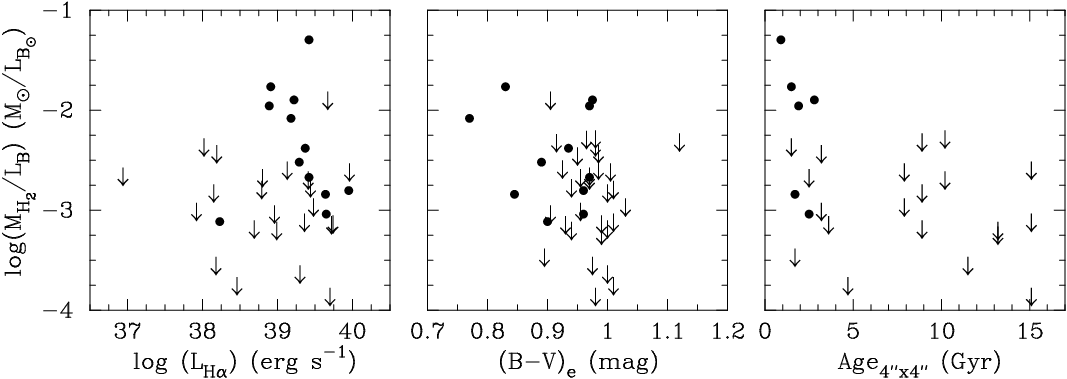}
\caption{Molecular gas content (normalised by the total blue
  luminosity) versus various star formation indicators for the {\tt
  SAURON} representative sample. {\bf Left:} H$\alpha$ luminosity
  within the {\tt SAURON} field-of-view, derived assuming an
  H$\alpha/$H$\beta$ flux ratio of $2.86$ \citep{setal06}. {\bf
  Centre:} Effective $B-V$ colour (HyperLEDA; \citealt{zetal02}). {\bf
  Right:} Luminosity-weighted stellar age, derived from line-strength indices
  measured in a $4\arcsec\times4\arcsec$ central aperture assuming a
  single stellar population \citep{mcetal06}. Detections are shown
  with solid dots and upper limits by arrows.}
\label{fig:sf}
\end{figure*}

Figure~\ref{fig:sf} shows correlations of the molecular gas content
(normalised by the blue luminosity) with three star formation
indicators, namely the H$\alpha$ luminosity within the {\tt SAURON}
field-of-view \citep{setal06}, the effective $B-V$ colour and the age
of the stellar populations, the latter derived from line-strength
indices measured in a $4\arcsec\times4\arcsec$ central aperture
assuming a single stellar population \citep{mcetal06}. No clear
correlation is seen, except perhaps with colour, but it is clear that
CO-rich galaxies have preferentially large H$\alpha$ luminosities,
blue colours, and young ages. In fact, all the galaxies detected in CO
and with a measured age have a central luminosity-weighted stellar ages less
than $3.5$~Gyr. No clear trend is seen with the derived metallicity or
$\alpha$-element abundance (\citealt{mcetal06}; not shown), strongly
suggesting that the trends with line-strength indices discussed above
are in fact primarily or perhaps exclusively driven by age, with
little dependence on the metallicity or $\alpha$-element abundance.
%
%
%
%
%
%
%
%
\subsection{Correlations with stellar kinematics}
\label{sec:kin_corr}
It is interesting to enquire about possible correlations between the
CO content of the {\tt SAURON} sample galaxies and their stellar
dynamics. Using the stellar kinematic maps of \citet{eetal04} and
\citet{mcetal06}, it is
relatively easy to identify discrete kinematic structures such as
central stellar discs and KDCs. 
In fact, \citet{mcetal06} identified two classes of KDCs among the
{\tt SAURON} galaxies.  One class is made up of compact KDCs with diameters
$\la$ 300 pc, which are often (though not exclusively) young, and
the other class consists of 
homogeneously old and extended kpc-scale KDCs. Interestingly,
half of the compact KDCs are detected in CO (NGC3032, NGC4150 and
NGC7457; the three youngest ones), but none of the large KDCs
are.  While some of the KDCs may well have formed through
dissipationless processes, the presence of molecular gas in the young,
compact KDCs convincingly argues that at least some of the
KDCs form through dissipative processes in which gas sinks to the
center of the galaxy and forms stars.

In addition to the compact KDCs, there are other cases which show
probable associations between the molecular gas and stellar
kinematic substructures.
For example, NGC~4526 is the most CO-rich galaxy in the {\tt SAURON}
sample, and it has a large dynamically cold stellar disk
\citep{eetal04} which shows young stellar ages in its H$\beta$
absorption line strengths \citep{ketal06}.  
The properties of NGC~4459 are similar though not as extreme.  
These cases are analogous to the compact, young KDCs in that the
molecular gas can probably be linked to the growth of a kinematically
decoupled stellar component.

Surprisingly, however, in a broader comparison of molecular gas content
to all types of kinematic substructure, we find that
there is no strong correlation between the existence of a decoupled
kinematic component and the presence of molecular gas. 
The fraction of galaxies with (or without) apparent kinematic
substructure in the maps of \citet{eetal04} is the same among CO detections and non-detections (and
among the entire {\tt SAURON} sample).  Conversely, the CO detection
(or non-detection) rate is the same among galaxies with and without
stellar kinematic substructure (and again among the entire {\tt
SAURON} sample).  The sample thus contains some galaxies with kinematic
substructure and no detectable molecular gas, as well as some galaxies
with molecular gas and no detectable kinematic substructure.

These facts do not lend themselves to a simple, straightforward picture
of the relationships between molecular gas and kinematic substructure.
In some cases there may be a direct link as a new stellar substructure
grows out of the molecular gas; in other cases the 
stellar kinematic substructures may be relics of past events completely
unrelated to the current molecular gas content. 
Synthesis
observations of our detections should shed new light on those
questions, allowing direct comparisons of the distribution and
kinematics of CO, stars, and ionised gas at similar spatial
resolutions.

%
%
\subsection{Aperture correction}
\label{sec:ap_corr}
As mentioned above, the 30~m beam at both CO(1-0) and CO(2-1) only
covers a small fraction of the optical extent of the
galaxies. Although the results of \citet{y02,y05} suggest that our
beams should recover all the CO in typical nearby early-type galaxies,
the same could in principle also be true of the molecular gas extent,
and one should correct the CO fluxes to a fixed physical aperture for
all galaxies in the sample. That aperture should preferentially be
related to the molecular gas extent, but failing this (since we do not
know the typical CO distribution), to a characteristic optical radius.

We have thus attempted to correct our fluxes assuming a uniform
molecular gas surface density within one optical effective radius
$R_{\rm e}$, within the optical radius $R_{25}$,
or assuming an exponential
distribution of effective radius equal to that in the optical. 
Such corrections were often large and only give the general impression
of adding even more scatter to the plots we have shown.
This suggests
that the molecular gas does not have a characteristic spatial extent and is
not strongly correlated to the stellar extent.
%
%
\section{FAR-INFRARED-RADIO CORRELATION}
\label{sec:fir-radio}
As illustrated above, galaxies which contain substantial amounts of
molecular gas are likely to have some star formation activity. As a
further attempt to understand the star formation properties of the
{\tt SAURON} early-type galaxies, we turn to an analysis of their FIR
and radio continuum emission. 
It has
been known for some time that if one defines a logarithmic 
FIR to radio continuum flux ratio as
\begin{equation}
q\equiv\log\biggl(\frac{FIR}{3.75\times10^{12}~{\rm W}~{\rm m}^{-2}}\biggr) - 
       \log\biggl(\frac{S_{1.4 {\rm GHz}}}{{\rm W}~{\rm m}^{-2}~{\rm Hz}^{-1}}\biggr)\,,
\end{equation}
then the bulk of the gas-rich, star forming galaxies lie in a tight
band at $q\approx2.34$, with a dispersion of about $0.26$~dex
\citep*{c92,yrc01}. The small dispersion in $q$ values is normally
interpreted to mean that both the FIR emission and the cm-wave radio
continuum emission are powered by star formation activity. Galaxies
whose cm-wave radio continuum emission is dominated by an active
galactic nucleus (AGN) are usually easy to identify by their low $q$
values.

There is good evidence that the radio continuum emission of early type
galaxies arises from both star formation and AGN activity, different
processes dominating in different galaxies. From an early analysis of
the NVSS data \citep[e.g.][]{ccgyptb98}, \citet{cc98} found that one
quarter of the E and S0 galaxies detected by NVSS and {\it IRAS}
``have IR/radio flux density ratios suggestive of star formation,''
whereas the other three quarters are probably powered by
AGN. \citet{wh88} obtained high resolution radio images and showed
that six early-type galaxies which lie on the FIR-radio correlation
indeed have extended kpc-scale radio morphologies, consistent with
star formation rather nuclear activity.

If the early-type galaxies which lie on the FIR-radio correlation are
indeed experiencing star formation, and the star formation has an
efficiency typical of that in spirals, we should expect to find that
the early-type galaxies on the FIR-radio correlation are also the ones
with molecular gas. In short, a comparison of $q$ values and CO
content in early-type galaxies should reveal something about the
nature of star formation and the fate of the molecular gas in those
objects.

\begin{figure}
\includegraphics[width=8cm]{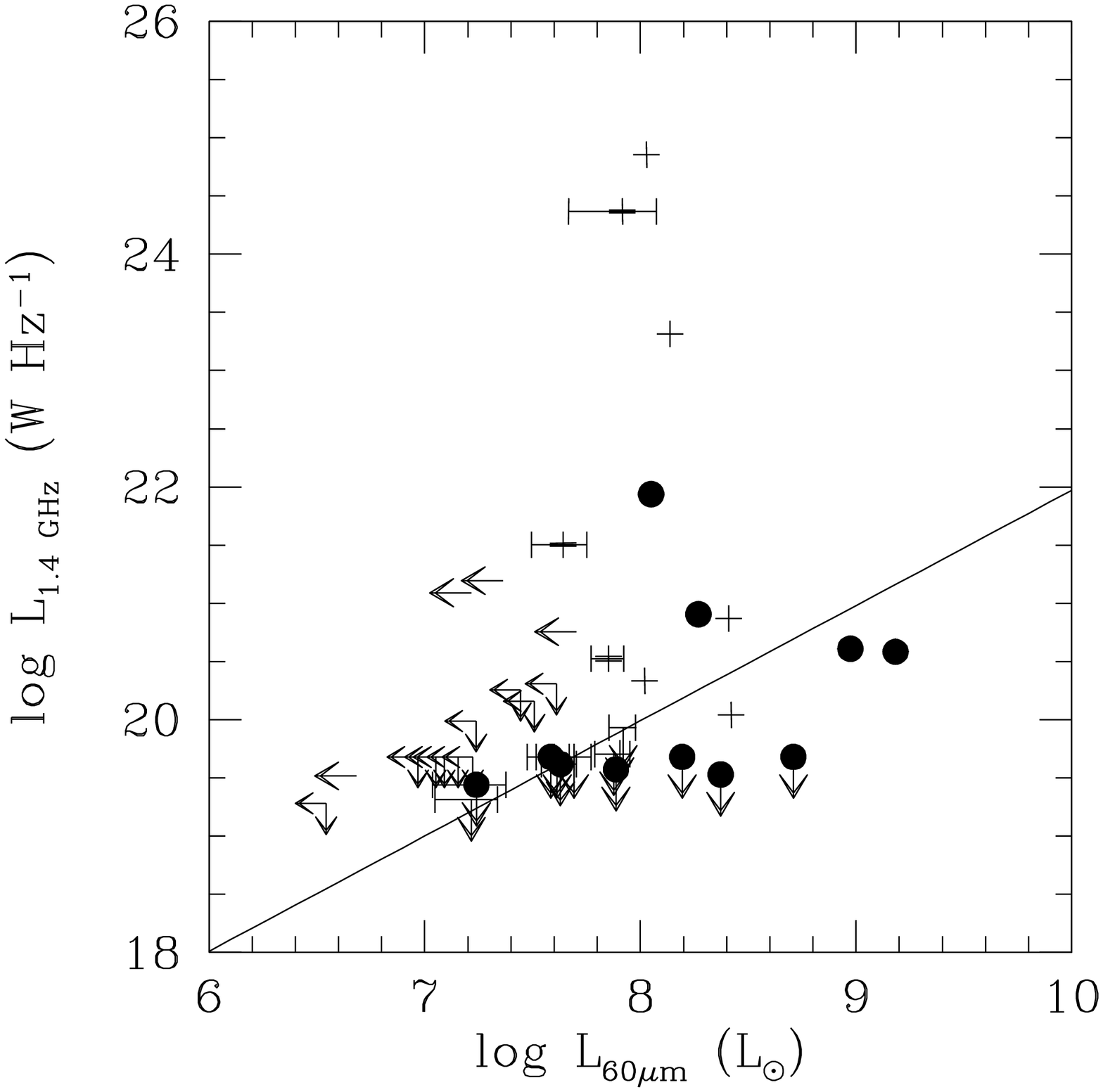}
\caption{$1.4$~GHz radio luminosities versus {\it IRAS} $60$~$\micron$
luminosities. CO detections are indicated with filled circles and
upper limits by arrows. Error bars ($1\sigma$) are shown only if they
are larger than the symbol size. The solid line shows the fit
$\log(L_{1.4 {\rm GHz}})=0.99\log(L_{60\micron}/L_{\sun})+12.07$ which
\citet{yrc01} obtained for a sample of $1809$ galaxies of all types
with $S_{60\micron}>2$~Jy. Distance uncertainties are not incorporated
into the error bars.}
\label{fig:1.4vs60}
\end{figure}

\begin{figure}
\includegraphics[width=8cm]{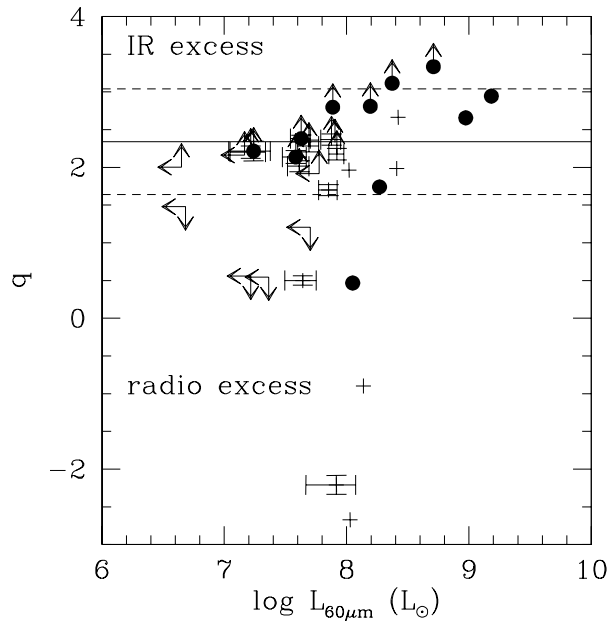}
\caption{{\tt IRAS} FIR/radio continuum flux ratio $q$ (see text)
versus $60$~$\micron$ luminosities. The solid line marks the average
value $q=2.34$ derived by \citet{yrc01}. The dotted lines delineate
the ``radio-excess'' and ``IR-excess'' regions defined by
\citet{yrc01} as having FIR/radio flux ratios $5$ times larger or
smaller than the mean. For a dispersion of $0.26$~dex, these lines lie
$\approx2.7\sigma$ away from the mean. As in Figure~\ref{fig:1.4vs60},
CO detections are indicated with filled circles, upper limits by
arrows, and error bars are shown only if they are larger than the
symbol sizes.}
\label{fig:q}
\end{figure}

Figures \ref{fig:1.4vs60}--\ref{fig:q} and Table \ref{tab:qfirad} show
FIR and radio continuum fluxes, luminosities and $q$ values for the
sample galaxies. {\it IRAS} fluxes are taken from the NASA
Extragalactic
Database\footnote{http://nedwww.ipac.caltech.edu/index.html} (NED).
The $1.4$~GHz fluxes were obtained from the NVSS catalogue
\citep{ccgyptb98} with three exceptions: NGC4261, NGC4374 and NGC4486
are highly resolved by NVSS so their fluxes are taken from
\citet{wb92}. Galaxies that are not detected in the NVSS are assigned
a $3\sigma$ upper limit of $1.5$~mJy. Of the $47$ galaxies with CO
data, we are able to calculate a $q$ value or limit for $32$
($68\%$). The remaining $15$ are undetected at all three wavelengths
or were not observed by {\it IRAS}.
%
%

\begin{table*}
\caption{FIR and radio parameters of the sample galaxies.}
\label{tab:qfirad}
\begin{tabular}{lrrrrr}
\hline
Galaxy & D & $S_{100\micron}$\phantom{00} & $S_{60\micron}$\phantom{000} & $P_{20{\rm cm}}$\phantom{00} & $q$\phantom{00000} \\
       & (Mpc) &(Jy)\phantom{0000}             & (Jy)\phantom{00000}            & (mJy)\phantom{00}            & \\
(1)    & (2) & (3)\phantom{00000}              & (4)\phantom{000000}             & (5)\phantom{0000}              & (6)\phantom{0000} \\ 
\hline
NGC 474 & 31.62 & $<$ 0.10\phantom{0} \phantom{(0.00)} & $<$ 0.027   \phantom{(0.00)} & $<$   1.5 \phantom{(00.0)} &     --- \phantom{00(0.00)} \\
NGC 524 &32.81&     2.05\phantom{0} (0.13)           &     0.76\phantom{0} (0.03)           &       3.1 (\phantom{0}0.4) &     2.64 (0.06\phantom{0}) \\
NGC 821 &23.55&     0.50\phantom{0} (0.12)           & $<$ 0.041           \phantom{(0.00)} & $<$   1.5 \phantom{(00.0)} & $>$ 2.13 \phantom{(0.000)} \\
NGC1023  & 10.28& $<$ 0.075           \phantom{(0.00)} & $<$ 0.032           \phantom{(0.00)} & $<$   1.5 \phantom{(00.0)} &     --- \phantom{00(0.00)} \\
NGC2549  & 16.75&     0.37\phantom{0} (0.13)           &     0.26\phantom{0} (0.05)           & $<$   1.5 \phantom{(00.0)} & $>$ 2.37 \phantom{(0.000)} \\
NGC2679$^{\rm a}$ &30.62&     --- \phantom{00(0.00)}           &     --- \phantom{00(0.00)}           & $<$   1.5 \phantom{(00.0)} &     --- \phantom{00(0.00)} \\
NGC2685 &14.39& 1.87\phantom{0} (0.11)           &     0.36\phantom{0} (0.04)           & $<$   1.5 \phantom{(00.0)} & $>$ 2.80 \phantom{(0.000)} \\
NGC2695 &23.23& $<$ 0.11\phantom{0} \phantom{(0.00)} & $<$ 0.031   \phantom{(0.00)} & $<$   1.5 \phantom{(00.0)} &     --- \phantom{00(0.00)} \\
NGC2699 &23.23&     --- \phantom{00(0.00)}           &     --- \phantom{00(0.00)}           & $<$   1.5 \phantom{(00.0)} &     --- \phantom{00(0.00)} \\
NGC2768 &21.48&  1.37\phantom{0} (0.06)           &     0.39\phantom{0} (0.03)           &      14.5 (\phantom{0}0.6) &     1.74 (0.03\phantom{0}) \\
NGC2974 &24.32&     1.90\phantom{0} (0.05)           &     0.42\phantom{0} (0.03)           &      10.4 (\phantom{0}0.5) &     1.98 (0.03\phantom{0}) \\
NGC3032 &21.68&     4.7\phantom{00} (0.09)           &     1.94\phantom{0} (0.03)           &       7.2 (\phantom{0}0.5) &     2.66 (0.03\phantom{0}) \\
NGC3156 &15.14&     0.61\phantom{0} (0.08)           &     0.18\phantom{0} (0.03)           & $<$   1.5 \phantom{(00.0)} & $>$ 2.38 \phantom{(0.000)} \\
NGC3377 &10.67&     0.35\phantom{0} (0.07)           &     0.14\phantom{0} (0.05)           & $<$   1.5 \phantom{(00.0)} & $>$ 2.20 \phantom{(0.000)} \\
NGC3379 &10.67& $<$ 0.11\phantom{0} \phantom{(0.00)} & $<$ 0.041   \phantom{(0.00)} &       2.4 (\phantom{0}0.5) & $<$ 1.48 \phantom{(0.000)} \\
NGC3384 &10.67&  0.45\phantom{0} (0.09)           & $<$ 0.038           \phantom{(0.00)} & $<$   1.5 \phantom{(00.0)} & $>$ 2.00 \phantom{(0.000)} \\
NGC3414 &20.14&     0.56\phantom{0} (0.19)           &     0.25\phantom{0} (0.03)           &       4.4 (\phantom{0}0.4) &     1.96 (0.08\phantom{0}) \\
NGC3489  &10.67 &     --- \phantom{00(0.00)}           &     --- \phantom{00(0.00)}           & $<$   1.5 \phantom{(00.0)} &     --- \phantom{00(0.00)} \\
NGC3608  &15.56 &     --- \phantom{00(0.00)}           &     --- \phantom{00(0.00)}           & $<$   1.5 \phantom{(00.0)} &     --- \phantom{00(0.00)} \\
NGC4150  &13.68 &     2.67\phantom{0} (0.06)           &     1.22\phantom{0} (0.04)           & $<$   1.5 \phantom{(00.0)} & $>$ 3.12 \phantom{(0.000)} \\
NGC4261$^{\rm a}$ &31.62&     0.15\phantom{0} (0.05)  &     0.08\phantom{0} (0.04)           &      19 (0.6) $\times10^3$ & $  -2.21$ (0.13\phantom{0}) \\
NGC4262 &16.29&  0.39\phantom{0} (0.12)           &     0.18\phantom{0} (0.04)           & $<$   1.5 \phantom{(00.0)} & $>$ 2.28 \phantom{(0.000)} \\
NGC4270 &16.29& $<$ 0.094           \phantom{(0.00)} & $<$ 0.041           \phantom{(0.00)} & $<$   1.5 \phantom{(00.0)} &     --- \phantom{00(0.00)} \\
NGC4278 &13.68&   1.86\phantom{0} (0.06)           &     0.58\phantom{0} (0.05)           &     385\phantom{.0} (12\phantom{.0}) & 0.47 (0.02\phantom{0}) \\
NGC4365$^{\rm a}$& 17.95&     0.65\phantom{0} (0.13)      & $<$ 0.044   \phantom{(0.00)} & $<$   1.5 \phantom{(00.0)} & $>$ 2.16 \phantom{(0.000)} \\
NGC4374 &16.29&   1.16\phantom{0} (0.12)           &     0.5\phantom{00} (0.03)           &     6.5 (0.2) $\times10^3$ & $ -0.90$ (0.03\phantom{0}) \\
NGC4382 &16.29& $<$ 0.076           \phantom{(0.00)} &     0.15\phantom{0} (0.03)           & $<$   1.5 \phantom{(00.0)} & $>$ 1.94 \phantom{(0.000)} \\
NGC4387 &16.29& $<$ 0.178           \phantom{(0.00)} & $<$ 0.052           \phantom{(0.00)} & $<$   1.5 \phantom{(00.0)} &     --- \phantom{00(0.00)} \\
NGC4458 &16.29& $<$ 0.142           \phantom{(0.00)} & $<$ 0.034           \phantom{(0.00)} & $<$   1.5 \phantom{(00.0)} &     --- \phantom{00(0.00)} \\
NGC4459 &16.29&     4.82\phantom{0} (0.13)           &     1.87\phantom{0} (0.07)           & $<$   1.5 \phantom{(00.0)} & $>$ 3.34 \phantom{(0.000)} \\
NGC4473 &16.29& $<$ 0.107           \phantom{(0.00)} & $<$ 0.061           \phantom{(0.00)} & $<$   1.5 \phantom{(00.0)} &     --- \phantom{00(0.00)} \\
NGC4477 &16.29&     1.41\phantom{0} (0.10)           &     0.57\phantom{0} (0.05)           & $<$   1.5 \phantom{(00.0)} & $>$ 2.81 \phantom{(0.000)} \\
NGC4486 &16.29&     0.41\phantom{0} (0.10)           &     0.39\phantom{0} (0.04)           &     224 (6.7) $\times10^3$ & $  -2.67$ (0.047)           \\
NGC4526 &16.29&    17.1\phantom{00} (0.09)           &     5.56\phantom{0} (0.05)           &      12\phantom{.0} (\phantom{0}0.5) & 2.94 (0.02\phantom{0}) \\
NGC4546 &16.29&     0.89\phantom{0} (0.22)           &     0.26\phantom{0} (0.05)           &      10.5 (\phantom{0}0.5) &     1.70 (0.07\phantom{0}) \\
NGC4550 &16.29&     0.25\phantom{0} (0.09)           &     0.14\phantom{0} (0.03)           & $<$   1.5 \phantom{(00.0)} & $>$ 2.14 \phantom{(0.000)} \\
NGC4552 &16.29&  0.53\phantom{0} (0.06)     &     0.16\phantom{0} (0.05)   &  100\phantom{.0} (\phantom{0}3\phantom{.0}) & 0.50 (0.06\phantom{0}) \\
NGC4564 & 16.29& $<$ 0.19\phantom{0} \phantom{(0.00)} & $<$ 0.060    \phantom{(0.00)} & $<$   1.5 \phantom{(00.0)} &     --- \phantom{00(0.00)} \\
NGC4570  &16.29& $<$ 0.11\phantom{0} \phantom{(0.00)} & $<$ 0.045   \phantom{(0.00)} & $<$   1.5 \phantom{(00.0)} &     --- \phantom{00(0.00)} \\
NGC4621  &16.29& $<$ 0.094           \phantom{(0.00)} & $<$ 0.050           \phantom{(0.00)} & $<$   1.5 \phantom{(00.0)} &     --- \phantom{00(0.00)} \\
NGC4660  &16.29& $<$ 0.10\phantom{0} \phantom{(0.00)} & $<$ 0.048   \phantom{(0.00)} & $<$   1.5 \phantom{(00.0)} &     --- \phantom{00(0.00)} \\
NGC5198  &36.31& $<$ 0.077           \phantom{(0.00)} & $<$ 0.037           \phantom{(0.00)} &       3.6 (\phantom{0}0.4) & $<$ 1.21 \phantom{(0.000)} \\
NGC5308  &28.31& $<$ 0.089           \phantom{(0.00)} & $<$ 0.039           \phantom{(0.00)} & $<$   1.5 \phantom{(00.0)} &     --- \phantom{00(0.00)} \\
NGC5813 &26.30& $<$ 0.10\phantom{0} \phantom{(0.00)} & $<$ 0.023 \phantom{(0.00)} & 14.8 (\phantom{0}1\phantom{.0}) & $<$ 0.56 \phantom{(0.000)} \\
NGC5831  &22.80&     --- \phantom{00(0.00)}           &     --- \phantom{00(0.00)}           & $<$   1.5 \phantom{(00.0)} &     --- \phantom{00(0.00)} \\
NGC5838  &18.71&     1.67\phantom{0} (0.09)           &     0.73\phantom{0} (0.04)           &       2.6 (\phantom{0}0.4) &     2.66 (0.07\phantom{0}) \\
NGC5845  &21.78&     0.23\phantom{0} (0.10)           &     0.17\phantom{0} (0.02)           & $<$   1.5 \phantom{(00.0)} & $>$ 2.18 \phantom{(0.000)} \\
NGC5846  &24.89& $<$ 0.127   \phantom{(0.00)} & $<$ 0.036  \phantom{(0.00)} &      21\phantom{.0} (\phantom{0}1.3) & $<$ 0.55 \phantom{(0.000)} \\
NGC5982 &41.88&     0.37\phantom{0} (0.04)           & $<$ 0.033           \phantom{(0.00)} & $<$   1.5 \phantom{(00.0)} & $>$ 1.92 \phantom{(0.000)} \\
NGC6548$^{\rm a}$ &33.57& $<$ 0.119   \phantom{(0.00)} & $<$ 0.035    \phantom{(0.00)} & $<$   1.5 \phantom{(00.0)} &     --- \phantom{00(0.00)} \\
NGC7332  & 19.23 &     0.41\phantom{0} (0.11)           &     0.21\phantom{0} (0.03)           & $<$   1.5 \phantom{(00.0)} & $>$ 2.33 \phantom{(0.000)} \\
NGC7457  & 12.36&     0.45\phantom{0} (0.19)           &     0.11\phantom{0} (0.04)           & $<$   1.5 \phantom{(00.0)} & $>$ 2.22 \phantom{(0.000)} \\
\hline
\end{tabular}
\\
Columns: (1) Source name; (2) adopted distance; (3)--(4) $100$ and $60$~$\micron$ fluxes
(NED); (5) $20$~cm radio continuum power \citep{ccgyptb98,wb92}; (6)
FIR/radio continuum flux ratio. Uncertainties are quoted in
parentheses.\\
Notes: a: galaxies observed by {\tt SAURON} but not part of the
representative sample of $48$ E/S0s. 
\end{table*}
%
%

%
%

The {\tt SAURON} sample of early-type galaxies spans the full range of
FIR and radio properties. Following \citet{yrc01}, we define
``radio-excess'' and ``IR-excess'' galaxies to have $q$ values
respectively higher and lower than the average by $0.7$~dex
($\approx2.7\sigma$). Of the $32$ galaxies for which we have both CO
data and a $q$ value, $9$ galaxies\footnote{NGC3379, NGC4261, NGC4278,
NGC4374, NGC4486, NGC4552, NGC5198, NGC5813 and NGC5846 are
radio-excess galaxies.} fall in the radio-excess category. Their radio
continuum emission is most likely dominated by AGN activity
\citep{ry04}. Two galaxies fall in the IR-excess
category\footnote{NGC4150 and NGC4459 are IR-excess galaxies.}, which
is more poorly understood. As discussed more fully by \citet{yrc01},
an IR excess may indicate unusually warm dust from a compact starburst
or a dust-enshrouded AGN. Alternatively, an IR excess could suggest
that the galaxy has a smaller-than-usual synchrotron output due to a
weak interstellar magnetic field. Such IR excesses are rare, however,
only accounting for $9$ of $1809$ IR-selected galaxies in the sample
of \citet{yrc01}. In our sample, $21$ galaxies having both CO data and
a $q$ value lie within the ``average'' band of $q=2.34\pm0.7$,
although $14$ of those only have upper limits for their radio
continuum flux, and could conceivably move into the IR-excess category
if their radio continuum fluxes are a good deal smaller than the
current $1.5$~mJy limit.

The CO detection rate is significantly lower among the radio-excess
galaxies ($1/9$ or $11\%$) than among the ``average'' and IR-excess
galaxies ($10/23$ or $43\%$) (both IR-excess galaxies are detected).
Thus, early-type galaxies with $q\ga2.3$ are more likely to also
contain molecular gas, the raw material for star formation. This
association between molecular gas and $q\ga2.3$ bolsters the
contention of \citet{wh88} that early-type galaxies with a $q$ value
similar to that of spirals are also experiencing star formation.

If it is true that early-type galaxies with $q\ga2.3$ have FIR and
radio emissions which are primarily powered by star formation, then
$23$ of the $32$ {\tt SAURON} early-type galaxies with CO data and
estimated $q$ values show evidence for star formation
activity. Curiously, however, several galaxies which are not detected
in CO emission also have $q\approx2.3$. Especially notable are
NGC2974, NGC3414, NGC4546 and NGC5838, which have FIR and radio fluxes
within the ranges occupied by galaxies detected in CO (e.g.\
Fig.~\ref{fig:1.4vs60}). On the other hand, NGC2974 does show evidence
for recent star formation in the ultraviolet \citep{jbykd07}, and all
four galaxies have substantial amounts of ionised gas
\citep[e.g.][]{setal06}. If $q\approx2.3$ really is an indicator of
star formation activity, then perhaps these galaxies have abundant
molecular gas but low CO/H$_2$ ratios, or alternatively they could
have smaller molecular gas contents but more efficient star
formation. There might even also be some other processes beside star
formation which produce their FIR and radio fluxes.

The entire sample having CO data contains $22$ galaxies traditionally
classified as elliptical and $25$ lenticulars. As discussed in
Section~\ref{sec:incidence}, CO is detected in only
$2$ ellipticals but in $10$ lenticulars, a much higher detection
rate. In contrast, the ellipticals show more AGN activity. Of all
galaxies for which we could calculate $q$ (see
Table~\ref{tab:qfirad}), $9$ of $17$ ellipticals show a radio excess
but none of the $17$ S0s does. We caution however that detailed
studies of the morphology and kinematics of the {\tt SAURON}
early-type galaxies \citep[e.g.][]{eetal04,setal06} prompt some
reclassifications.
%
%
\section{CONCLUSIONS}
\label{sec:summary}
We surveyed in the CO(1-0) and CO(2-1) lines early-type galaxies
observed in the optical with the {\tt SAURON} integral-field
spectrograph. Taking into account literature results, the detection
rate for the {\tt SAURON} representative sample of $48$ E/S0s is
12/43 or $28\%$. This is lower than previous surveys but differences can
probably be explained by differences between the samples. Our data
extend the usual correlations between FIR luminosity, dust mass and
molecular mass of later galaxy types, as well as the related
correlation between star formation and molecular gas surface
density. Unsurprisingly, the more massive and luminous earlier type
galaxies have a relatively lower molecular gas content and the local
galaxy density does not appear to have a strong effect on the
molecular content.

Although weak, indicative trends are nevertheless observed with
various optical absorption line-strength indices. CO-rich galaxies tend
to have higher H$\beta$ and lower Fe5015 and Mg$b$ indices. CO-rich
galaxies also show the strongest evidence of star formation, probed in
a variety of ways, suggesting that those trends are primarily driven
by stellar age. Analysing the FIR and radio fluxes, we similarly see
that so-called ``radio-excess'' galaxies, presumably powered by active
galactic nuclei, are significantly poorer in molecular gas than
so-called ``IR-excess'' galaxies, presumably powered by star
formation. These trends support the recent {\it Galaxy Evolution
Explorer} (GALEX) ultraviolet results (among others), which suggest
that a significant number of early-type galaxies may be undergoing
a small amount of star formation activity
\citep[e.g.][]{yetal05}. Here, however, we have apparently found the
raw material for this residual star formation, and have demonstrated
that it is actively being transformed. This might not be surprising in
spiral galaxies, but it is a significant step for early-types.

Overall, however, the weakness of the correlations between the CO
content and most optically-derived parameters suggests that, in many
of these galaxies, the molecular gas does not
come from internal sources such as stellar mass loss.   Instead the gas has
probably been accreted and has properties largely independent from the
old stellar component.

The current CO detections offer the possibility to pursue those issues
with much improved spatially-resolved synthesis data, which can be
directly compared to optical integral-field observations at similar
spatial resolution. Much work along those lines is already ongoing.
%
%
\section*{Acknowledgments}
The authors would like to thank the many people from the {\tt SAURON}
team, and in particular Eric Emsellem, for useful discussions. FC and
LMY acknowledge support from the PPARC visitor grant
PPA/V/S/2002/00553 to Oxford University during part of this work. LMY
also thanks the Oxford Astrophysics Department for its hospitality
during sabbatical work. This work is based on observations made with
the IRAM 30~m telescope, in Pico Veleta, near Granada, Spain. This
project made use of the HyperLEDA and NED databases.
%
%

%
\end{document}